\documentclass[acmsmall,screen,nonacm, review=false]{acmart}

\AtBeginDocument{%
  \providecommand\BibTeX{{%
    \normalfont B\kern-0.5em{\scshape i\kern-0.25em b}\kern-0.8em\TeX}}}

\begin{document}

\title{Phantom: A High-Performance Computational Core for Sparse Convolutional Neural Networks}

\author{Mahmood Azhar Qureshi}
\email{mahmood102@ksu.edu}
\author{Arslan Munir}
\authornotemark[1]
\email{amunir@ksu.edu}
\affiliation{%
  \institution{Department of Computer Science, Kansas State University}
%  \streetaddress{P.O. Box 1212}
  \city{Manhattan}
  \state{Kansas}
  \country{USA}
  \postcode{66506}
}


\begin{abstract}
  Sparse convolutional neural networks (CNNs) have gained significant traction over the past few years as sparse CNNs can drastically decrease the model size and computations, if exploited befittingly, as compared to their dense counterparts. Sparse CNNs often introduce variations in the layer shapes and sizes, which can prevent dense accelerators from performing well on sparse CNN models. Recently proposed sparse accelerators like SCNN, Eyeriss v2, and SparTen, actively exploit the \textit{two-sided} or \textit{full} sparsity, that is, sparsity in both weights and activations, for performance gains. These accelerators, however, either have inefficient micro-architecture (Eyeriss v2, SCNN), which limits their performance, have no support for non-unit stride convolutions (SCNN) and fully-connected (FC) layers (SCNN, SparTen), or suffer massively from \textit{systematic load imbalance} (SCNN, Eyeriss v2). %It is also worth noting that some accelerators like SCNN perform worse under low sparsity (i.e. dense models). 
To circumvent these issues and support both sparse and dense models, we propose \textit{Phantom}, a multi-threaded, dynamic, and flexible neural computational core. \textit{Phantom} uses sparse binary mask representation to actively \textit{lookahead} into sparse computations, and dynamically schedule its computational threads to maximize the thread utilization and throughput. We also generate a two-dimensional (2D) mesh architecture of \textit{Phantom} neural computational cores, which we refer to as \textit{Phantom-2D} accelerator, and propose a novel dataflow that supports \textbf{all} layers of a CNN, including unit and non-unit stride convolutions, and FC layers. In addition, \textit{Phantom-2D} uses a two-level load balancing strategy to minimize the computational idling, thereby, further improving the hardware utilization. To show support for different types of layers, we evaluate the performance of the \textit{Phantom} architecture on VGG16 and MobileNet. Our simulations show that the \textit{Phantom-2D} accelerator attains a performance gain of $12\times$, $4.1\times$, $1.98\times$, and $2.36\times$, over dense architectures, SCNN, SparTen, and Eyeriss v2, respectively.
\end{abstract}




\maketitle

%%%%%%%%%%%%%%%%%%%%%%%%%%%%%%%%%%%%%%%%%%%%%%%%%%%%%%%%%%%%%%%%
%\vspace{-2 mm}
\section{Introduction}
%\vspace{-1 mm}
%%%%%%%%%%%%%%%%%%%%%%%%%%%%%%%%%%%%%%%%%%%%%%%%%%%%%%%%%%%%%%%%

Neural nets have been around since the 1940s, however, the first practically applicable neural network, referred to as the LeNet \cite{lenet}, was proposed in 1989. This neural network was designed to solve the problem of digit recognition in hand-written numeric digits. It paved the way for the development of neural networks responsible for various applications related to digit recognition like an ATM. The slow growth and a little to no adoption of neural networks in the early days is mainly due to the massive computational requirements involved with their processing which limited their study to theoretical concepts. Over the past decade, there has been an exponential growth in the research on deep neural networks (DNNs) with many new high accuracy DNNs being deployed for various applications. This has only been possible because of two factors. The first factor is the advancements in the processing power of semiconductor devices and technological breakthroughs in computer architecture. Nowadays, computers have significantly higher computing capability. This enables the processing of a neural network within a reasonable time frame, something that was not achievable in the early days. The second factor is the availability of a large amount of training datasets. As neural networks learn over time, providing huge amounts of training data enables better accuracy. For example, Facebook receives close to a billion user images per day, whereas, YouTube has 300h of video uploaded every minute \cite{DNNtutorial}. This enables the service providers to train their neural networks for targeted ad campaigns bringing in billions of dollars of ad revenue. Apart from their use in social media platforms, DNNs are impacting many other domains and expect to make a huge impact. One of the domains where DNNs have contributed significantly is speech processing. Nowadays, many applications have been developed that use DNNs to perform real-time speech recognition with unprecedented levels of accuracy \cite{speech2,speech22,speech23}. Many technology companies are also using DNNs to perform language translation used in a wide variety of applications. Google, for example, uses Google's Neural Machine Translation system (GNMT) \cite{google} which uses recurrent neural networks (RNNs), a type of DNN, for their language translation applications. Autonomous driving has been one of the biggest technological breakthroughs in the auto industry since the invention of the internal combustion engine. It is not a coincidence that the self-driving boom came at the same time when high accuracy DNNs became increasingly popular. Companies like Tesla and Waymo are using various types of self-driving technology including visual feeds and Lidar for their self-driving solutions. One thing which is common in all these solutions is the use of DNNs for visual perception of the road conditions which is the main back-end technology used in advanced driver assistance systems (ADAS). Another crucial area where DNNs have become increasingly useful is medicine. Nowadays, doctors can use AI-assisted medical imagery to perform various surgeries. AI systems use DNNs in genomics to gather insights about genetic disorders like autism \cite{autism,autism2}. DNNs are also useful in the detection of various types of cancers like skin and brain cancer \cite{skin,brain}. The advent of AI has also challenged many traditional security approaches that were previously deemed sufficient. The rollout of 5G technology has caused a massive surge of IoT-based deployments which traditional security approaches are not able to keep up with. Physical unclonability approaches \cite{pufipa,pufrla,pufrake} were introduced to protect this massive deployment of IoTs against security attacks with minimum cost overheads. These approaches, however, were also unsuccessful in preventing AI-assisted attacks using DNNs \cite{pufattack,pufattack1}. Researchers have now been forced to upgrade the security threat models to incorporate AI-based attacks \cite{AIattack00,AIattack0}. Because of a massive increase in AI-assisted cyber-attacks on cloud and datacenters, corporations and governments have realized that the best way of defeating offensive AI attacks is by incorporating AI-based defensive strategies.\par

Overall, the use of DNNs in various applications has seen exponential growth over the past decade and this trend has been on the rise for the past many years. The massive increase in DNN deployments on the edge devices requires the development of efficient processing architectures to keep up with the computational requirements for successful DNN inference. \par

%Deep neural networks (DNNs) have greatly improved over the last decade resulting in widespread deployment for computer vision tasks \cite{alexnet,goingdeeper}, and natural language processing\cite{languageModel}. 
Convolutional neural networks (CNNs) for vision artificial intelligence (AI) applications have reached an unprecedented accuracy ever since the introduction of AlexNet \cite{alexnet} about a decade ago. Many of the previously proposed high accuracy CNNs \cite{alexnet,vgg16,resnet,googlenet} have tremendous amounts of computations owing to a large number of model parameters. These parameters and the associated massive number of computations generate exorbitant amounts of data in the form of partial sums (psums) and feature maps. This massive data, in addition to the model parameters, raise concerns in regards to both compute and memory bandwidth. In addition, it also raises concerns about energy consumption of a neural network accelerator (NNA), since, on-chip memory is not sufficient to store the entire model, which, in some cases, can be in the order of hundreds of MBs. To support this massive model size of a CNN, off-chip DRAM memories are generally employed. It has been shown that the energy cost per fetch for 32b coefficients in an off-chip LPDDR2 DRAM is about 640pJ, which is about $6400\times$ the energy cost of a 32b integer ADD operation\cite{energy}. The power dissipation resulting from just the DRAM accesses would be well beyond the limits of an embedded mobile device employing the NNA. \par

%Various techniques have been developed to address the compute and memory bandwidth issues of an NNA, running a CNN. \textit{Mobilenets} \cite{mobnetv1,mobnetv2} were developed to reduce the total number of computations by splitting a regular convolution operation into separable convolutions (depthwise and pointwise), without incurring a loss in accuracy. Another widely used approach for decreasing the model size is the reduction in precision of both weights and activations using various quantization strategies \cite{logquant1,logquant2,fixquant1}. This again does not result in a significant loss in accuracy and reduces the model size by a considerable amount. Hardware implementations like Envision\cite{envision}, NeuroMAX\cite{neuromax}, UNPU\cite{UNPU}, and Stripes\cite{stripes} show how reduced bit precision, and quantization, translates into increased throughput and savings in energy.

Modern CNNs owe their high accuracy to deep layers and the non-linearity in their design. Typically, non-linearity is added by incorporating activation functions, the most common being the rectified linear unit (ReLU) \cite{DNNtutorial}. The ReLU converts all negative values in a feature map to zeros. Since the output of one layer is the input to the next layer, many of the computations, within a layer, involve multiplication with zeros. These feature maps containing zeros are referred to as \textit{one-sided} sparse feature maps. The multiplications resulting from this one-sided sparsity wastes compute cycles and decreases the \textit{effective} throughput and hardware utilization, thus, reducing the overall performance of the accelerator. It also results in high energy cost as the transfer of zeros to/from off-chip memory is a wasted memory access. In order to reduce the computational and memory access volume, previous works \cite{cnvlutin,cambriconX,eyeriss} have exploited this one-sided sparsity and displayed some performance improvements.\par

Compression of DNN models was introduced the first time in \cite{deepcompression}. Han et al. \cite{deepcompression} iteratively pruned the connections based on parameter threshold, and performed retraining to retain accuracy. This process resulted in \textit{two-sided} sparsity, i.e., sparsity in both weights and activations, which led to approximately 9$\times$ model reduction for AlexNet, and 13$\times$ reduction for VGG-16. It also resulted in 4 - 9$\times$ \textit{effective} compute reduction (depending on the model). These gains, in theory, are very promising, however, designing an accelerator that leverages the two-sided sparsity is quite challenging because of the following reasons:\par
\textit{1) Inconsistency in Data Accesses}: Sparsity is generally exploited through gating of computation whenever a zero is read in either weight or activation data. This type of gating translates into energy savings but has no impact on throughput as the compute cycle is wasted and no effective work is performed during a \textit{zero} read. Complex read logic needs to be implemented to discard the zeros, and instead perform effective computations on non-zero data. Some previous approaches \cite{eyerissv2,eie} use compressed sparse column (CSC) 
%and/or compressed sparse column (CSC) 
format to represent sparse data. These formats have variable lengths and make \textit{``looking ahead''} difficult if both the weight and the activation sparsity is being considered. Other than that, developing the complex control and read logic to process these formats can be quite challenging.\par
%and they add high hardware overheads for the accelerators.

\textit{2) Load Imbalance and Under Utilization of the Processing Element (PE) Array}: Many accelerator architectures have a two-dimensional array of PEs that process the data in a consistent manner to maximize the hardware utilization. Different dataflows have been developed that efficiently schedule the data onto the PEs to maximize the throughput\cite{DNNtutorial}. Sparsity brings inconsistency in the scheduling of data, introducing workload imbalance. The subsection of PEs which are provided with more sparse data have idle times while those provided with dense data are fully active. This bounds the throughput of the accelerator to the most active PEs, and therefore, leads to the under utilization of the PE array.\par

Considering the aforementioned issues, many previously proposed accelerators attempt to strike a balance between hardware complexity and performance improvements. 
%Cnvlutin\cite{cnvlutin} does not avoid transfer of zeros and only skips cycles for activations. The latter, even though can improve energy efficiency, has no effect on throughput. Eyeriss\cite{eyeriss} only gates computations for sparse activations. 
Eyeriss v2 \cite{eyerissv2} attempts to address the two-sided sparsity by using CSC format for both the activations and weights. It, however, fails to address the systematic load imbalance introduced due to variations in the density of the sparse matrices. It also requires complex read logic embedded within a PE that drastically increases the area by $\sim$ 93\% when compared to the original Eyeriss \cite{eyeriss}. 
%Cambricon-X \cite{cambriconX} does not store activations in compressed format while Cambricon-S\cite{cambriconS} forces regularity by employing coarse grain pruning that affects accuracy. Even though it discards zeros during computation, it still retrieves and stores them. 
EIE \cite{eie} exploits the two-sided sparsity, albeit \textit{only} in fully-connected layers. EIE's performance is equivalent to one-sided sparsity as it discards zeros in the filter but wastes compute cycles due to being idle. Sparse CNN (SCNN) \cite{scnn} targets two-sided sparsity but suffers heavily from inefficient microarchitecture and systematic load imbalance as explained in \cite{sparten}. It, also, can not handle non-unit stride convolutions and FC layers. SparTen \cite{sparten} addresses the issues in previous architectures and attempts to strike a balance between data reuse and load imbalance. 
%To address the complexity associated with the CSC compression format, 
Instead of CSC format, SparTen, instead, uses sparse bit mask to represent the location of zeros and non-zero data values. SparTen, however, needs an offline load balancing strategy, which it refers to as \textit{Greedy Balancing}, to address the systematic load imbalance. 
%The balancing is performed based on the filter density using either a software-only approach (GB-S), or a software-hardware hybrid (GB-H). 
This form of balancing adds extra latency and complicates the synchronization of various compute threads. 
%SparTen also employs \textit{Permuter} and \textit{Output Collector Units} for the computation clusters to merge and/or accumulate the outputs from \textit{independently} running compute units. These circuits require rather complex hardware and the complexity grows exponentially as the number of compute units are increased. SparTen also does not provide the architecture and/or implementation details of these circuits.\par
In addition, SparTen, like SCNN, has no support for FC layers.\par
To address all the issues described above, we propose \textit{Phantom}, a flexible and high throughput neural computational core which promises high hardware utilization. The core works for both dense and sparse networks by dynamically mapping valid computations on the processing threads. The core also addresses the systematic load imbalance by using a two-level, dynamic, load balancing strategy. Unlike some of the previous works, the core can work on any input layer, be it CONV or FC, and supports any type of convolution (regular or separable). In summary, the main contributions of this work include:
%\vspace{-1.5 mm}
\begin{itemize}
    \item A \textit{multi-threaded} neural computational core architecture, called \textbf{Phantom}, designed to maximize the hardware utilization and throughput of CNN models. \textit{Phantom} exploits the sparsity in \textit{both} weights, and activations, simultaneously, by incorporating simple, yet powerful circuits like ``\textit{Lookahead Mask}'', ``\textit{Top-Down Selector}'' and ``\textit{Thread Mapper}''. These circuits, when used in conjunction, map the \textit{effective}, \textit{non-zero} computations onto an array of multiplier threads within a PE. The core also has the capability to skip huge number of non-essential computations (\textit{zero\textsubscript{w}} $\times$ \textit{zero\textsubscript{a}}, \textit{zero\textsubscript{w}} $\times$ \textit{non-zero\textsubscript{a}}, \textit{non-zero\textsubscript{w}} $\times$ \textit{zero\textsubscript{a}}), while simultaneously favoring essential computations (\textit{non-zero\textsubscript{w}} $\times$ \textit{non-zero\textsubscript{a}}), without wasting compute cycles (subscripts ``w'' and ``a'' here refer to weights and activations, respectively). This drastically improves the core's hardware utilization, consequently improving the throughput. %\vspace{-2.0 mm}
    \item Addressing the systematic load imbalance by using a two-tiered, on-the-fly, load balancing strategy. Unlike some previous approaches, this balancing does not require offline processing or modification of the CNN model. %\vspace{-2.0 mm}
    \item Generating a two-dimensional (2D) mesh architecture of \textit{Phantom} neural computational cores, which we refer to as \textbf{Phantom-2D} accelerator. Unlike some previous works that only support either CONV layers or FC layers, we show how \textit{Phantom-2D} supports different CONV types (unit and non-unit stride) and FC layers, in addition to supporting both sparse and dense CNN models. Simulations show that the \textit{Phantom-2D} accelerator has a performance gain of $12\times$, $4.1\times$, $1.98\times$, and $2.36\times$, over dense architectures, SCNN, SparTen, and Eyeriss v2, respectively, while retaining the energy efficiency of SparTen.
    %\item Showing a drastic improvement in performance when compared to some recently proposed two-sided sparse accelerators. Simulations show that the \textit{Phantom-2D} accelerator has a performance gain of $12\times$, $4.1\times$, $1.98\times$, and $2.36\times$, over a dense architecture, SCNN, SparTen, and Eyeriss v2, respectively, while retaining the energy efficiency of SparTen.
\end{itemize}

%End of Introduction
%%%%%%%%%%%%%%%%%%%%%%%%%%%%%%%%%%%%%%%%%%%%%%%%%%%%%%%%%%%%%%%%

%%%%%%%%%%%%%%%%%%%%%%%%%%%%%%%%%%%%%%%%%%%%%%%%%%%%%%%%%%%%%%%%
%\vspace{-3.5 mm}
\section{Related Work}
%\vspace{-1 mm}
%%%%%%%%%%%%%%%%%%%%%%%%%%%%%%%%%%%%%%%%%%%%%%%%%%%%%%%%%%%%%%%%
Many dense architectures have been proposed in the literature that optimize compute \cite{snowflake,jouppi2017indatacenter,fixquant1} and memory bandwidth \cite{ShiDianNao,PuDianNao} for CNN inferences. Quantization of weights and activations using log \cite{logquant1,logquant2} and linear \cite{fixquant1,gupta2015deep} techniques further reduce the memory footprint. This does not result in a significant loss in accuracy and reduces the CNN model size by a considerable amount. Hardware implementations like Envision\cite{envision}, VWA\cite{VWA}, NeuroMAX\cite{neuromax}, UNPU\cite{UNPU}, and Stripes\cite{stripes}, show how reduced bit precision, efficient dataflow, and quantization, translates into increased throughput and savings in energy. Efficient data reuse-based accelerators \cite{fusedCNN,eyeriss} maximize the data reuse within different layers to minimize the memory accesses, thereby, reducing energy consumption. Separable accelerators \cite{fpgaCNN1,fpgaCNN2} implement efficient hardware on FPGA for accelerating separable convolutions. These accelerators, however, cannot handle a vast majority of CNNs that employ regular convolutions and FC layers. Bit-serial accelerators \cite{laconic,bittactical} use booth encoding to suppress the use of zero bits, and, thereby, reduce the total computations. These schemes, however, transfer zeros to and from memory which incurs SRAM area and energy. CirCNN\cite{circnn} uses block circulant matrices for weights to improve the performance. It, however, utilizes complex hardware to perform the FFT operations, and also, does not capture full sparsity. In-memory accelerators \cite{puma,isaac} use simple analog logic to implement matrix multiplications within memory. These accelerators, however, cannot exploit sparsity as it requires complex ALU and buffering logic. Analog circuits also suffer heavily from noise and process variations which can drastically reduce the CNN accuracy.\par
Sparse architectures try to reduce the compute and data volume by exploiting the naturally occurring zeros in weights or activations (one-sided), or both weights and activations (two-sided). Cnvlutin\cite{cnvlutin} and Cambricon-X\cite{cambriconX} exploit one-sided sparsity of either weights or input maps but not both. Cnvlutin, also, does not avoid transfer of zeros and only skips cycles for activations. Cambricon-X does not store activations in compressed format while Cambricon-S\cite{cambriconS} forces regularity by employing coarse grain pruning that affects accuracy. Even though it discards zeros during computation, it still retrieves and stores them. Tensaurus \cite{tensaurus} accelerates dense and sparse tensor factorizations by introducing a new dataflow which they refer to as compressed interleaved sparse slice (CISS) dataflow. Tensaurus, however, is capable of supporting only one-sided sparsity. Some recent sparse GEMM (SpGEMM) accelerators \cite{matraptor,extensor,sigma,sparch,outerspace,sparsePE} target general sparse-matrix, sparse-matrix multiplications. Extensor \cite{extensor} and Sigma \cite{sigma} use output stationary (inner product) dataflow for sparse matrix multiplications. Inner product, however, is inefficient against highly sparse matrices because every element of the rows and columns must be traversed even though there are less effectual computations (non-zero $\times$ non-zero). This leads to a massive amount of wasted computations. SpArch \cite{sparch} and OuterSPACE \cite{outerspace} use input stationary (or outer-product) dataflow to avoid the inefficiencies associated with the inner-product dataflow. Outer-product, however, gives poor output reuse as the partial outputs generated are more in quantity than the final outputs which can cause significant memory traffic. 
%Finally, MatRaptor \cite{matraptor} uses Gustavson's Algorithm \cite{gusta} (or row-stationary) to address the issues associated with the inner and the outer-product dataflows. 
Finally, MatRaptor \cite{matraptor} introduces channel cyclic sparse row (C\textsuperscript{2}SR) dataflow for better reuse and memory efficiency. It is a modified version of the CSR format but requires complex encoding for output matrices. %\textit{Phantom} addresses the issues present in many recent accelerators.  %\par
Finally, SCNN, SparTen, and Eyeriss v2, exploit the full two-sided sparsity, but, as explained previously, suffer from either inefficient micro-architecture, no support for FC layers and non-unit stride convolutions, complex PE design to incorporate CSC compression format, or systematic load imbalance. \textit{Phantom} addresses all of these issues while also providing higher performance and energy efficiency.

\begin{figure}[ht]
\centering
\includegraphics[width=0.90\textwidth,keepaspectratio]{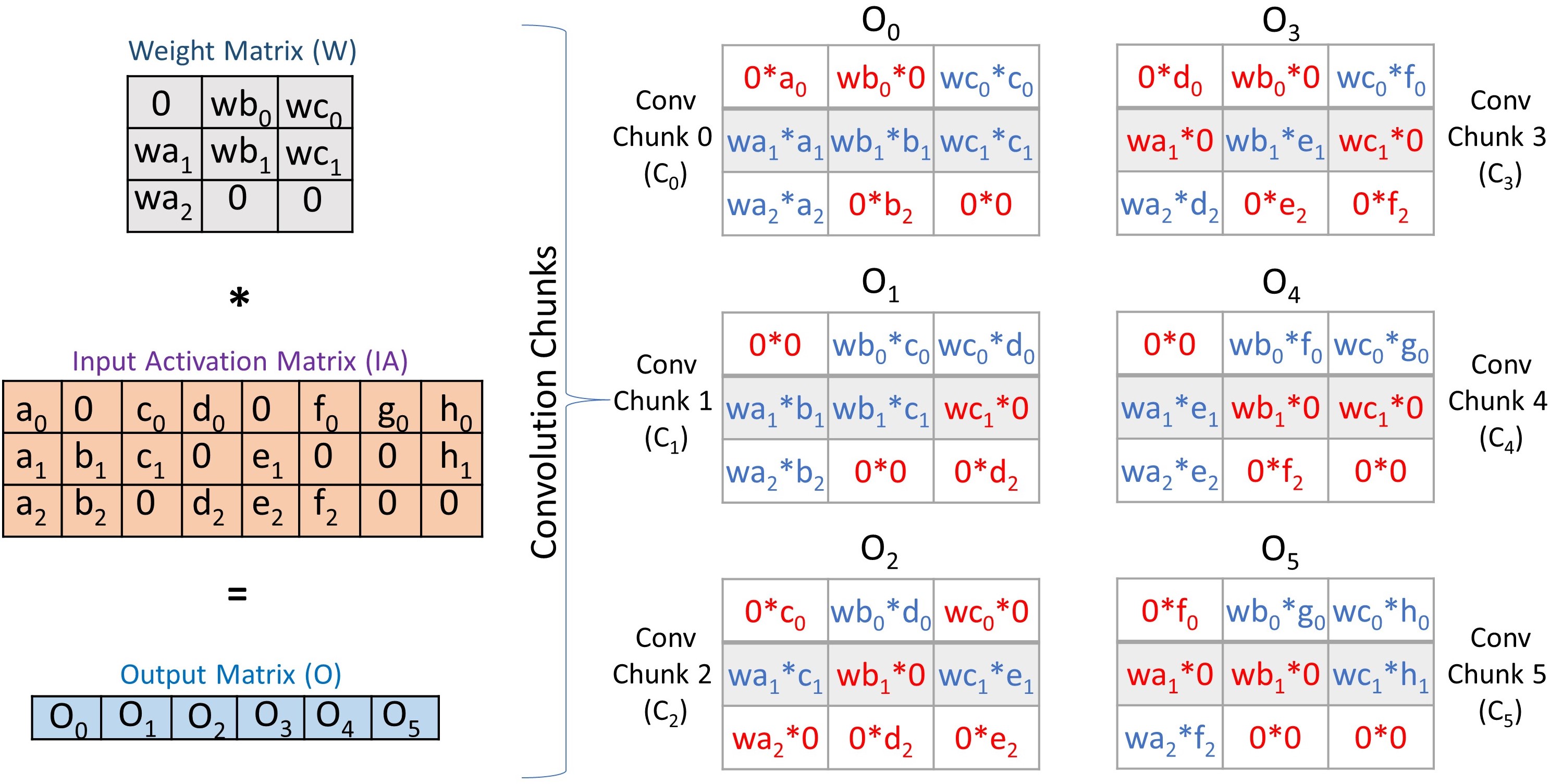}
\caption{$3\times 3$ convolution example}
\label{chunks}
\end{figure}

\begin{figure}[ht]
\centering
\includegraphics[width=0.90\textwidth,keepaspectratio]{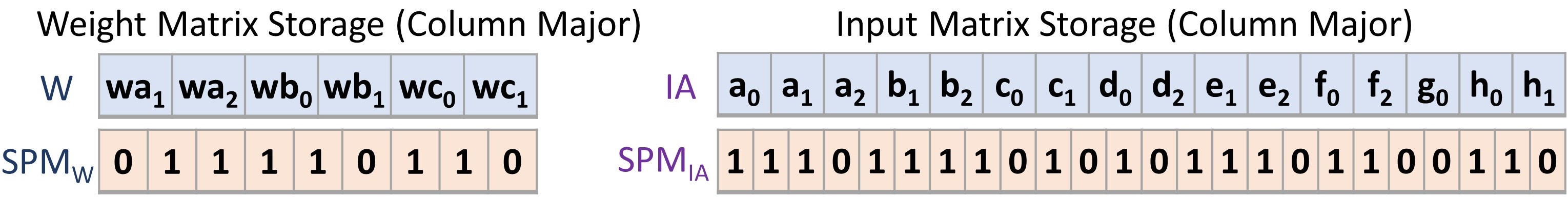}
\caption{Sparse mask representation}
\label{SPM_reps}
\end{figure}

%%%%%%%%%%%%%%%%%%%%%%%%%%%%%%%%%%%%%%%%%%%%%%%%%%%%%%%%%%%%%%%%
%\vspace{-4 mm}
\section{PHANTOM}
%\vspace{-1 mm}
%%%%%%%%%%%%%%%%%%%%%%%%%%%%%%%%%%%%%%%%%%%%%%%%%%%%%%%%%%%%%%%%
This section describes the architecture and inner workings of the \textit{Phantom} core. Recall from Section 1 that the \textit{Phantom} core in itself provides two major contributions. First, it considers both activation and weight sparsity simultaneously and \textit{looks ahead} into future computations to determine only the valid MAC operations (\textit{non-zero\textsubscript{w}} $\times$ \textit{non-zero\textsubscript{a}}). Second, because of the multi-threaded design of the PEs, the scheduling of data into each thread is handled dynamically in a non-linear fashion, based on the sparsity of input and weight matrices. This ensures that only valid computations are mapped on to the multi-threaded PEs and that the compute cycles are not wasted. This is opposed to the designs which schedule data into the PEs in a constant manner and gate the computations whenever zeros are read, thereby, wasting the compute cycles. \par
Figure \ref{chunks} shows a $3\times 3$ convolution example, where a $3\times 8$ input is convolved with a $3\times3$ filter to produce a $1\times 6$ output. The six individual convolution chunks which produce the output are also shown. The effective and ineffective multiplications are shown in blue and red, respectively. It can be seen that, on average, $55\%$ computations involve multiplication with zeros which results in wasted compute cycles. This causes a significant drop in the effective throughput, as many cycles are wasted in \textit{zero} multiplications. The \textit{Phantom} core addresses this issue and uses look-ahead masking to maximize the effective throughput by skipping ineffectual computations involving zeros.

%%%%%%%%%%%%%%%%%%%%%%%%%%%%%%%%%%%%%%%%%%%%%%%%%%%%%%%%%%%%%%%%

%%%%%%%%%%%%%%%%%%%%%%%%%%%%%%%%%%%%%%%%%%%%%%%%%%%%%%%%%%%%%%%%
\subsection{Sparse Mask Representation}
%%%%%%%%%%%%%%%%%%%%%%%%%%%%%%%%%%%%%%%%%%%%%%%%%%%%%%%%%%%%%%%%
%\begin{figure}
%\centering
%\includegraphics[width=0.80\textwidth,keepaspectratio]{SPM_rep.jpg}
%%\vspace{-4 mm}
%\caption{Sparse Mask Representation}
%\label{SPM_reps}
%%\vspace{-6 mm}
%\end{figure}

Unlike the recent approaches that use compressed sparse row (CSR) or CSC formats to represent non-zero data \cite{cambriconX,cnvlutin,eie}, we use a binary mask called \textit{sparse mask} for both weight and activation data. Sparse mask provides an efficient and simplistic way for data representation and enables identification of zero and non-zero data without explicitly storing zeros. It also does not require storage of \textit{count} and \textit{data pointers} which are needed for CSC (and CSR) formats. Figure \ref{SPM_reps} shows the equivalent sparse mask representation and storage of the input and the weight matrix shown in Figure \ref{chunks}. For a particular matrix, two arrays are stored in the column major format. The data array contains the non-zero data, whereas, the binary array contains the sparse mask. The ones in the sparse mask array represent the location of \textit{stored} non-zero values, whereas, zeros represent \textit{unstored} zero data. To process a specific type of convolution, the sparse mask and the data array is broken into chunks and scheduled to the core. 
Although, the core can work on any type of CONV or FC layers, for ease of understanding, we will explain the working of \textit{Phantom} core using the $3\times3 $ unit stride convolution example, shown in Figure \ref{chunks}. %The corresponding 6 chunks of length 9 ($3\times3$) sparse masks and the associated data vectors are also shown in Figure \ref{SPM_reps}.

%%%%%%%%%%%%%%%%%%%%%%%%%%%%%%%%%%%%%%%%%%%%%%%%%%%%%%%%%%%%%%%%

\begin{figure}[ht]
\centering
\includegraphics[width=0.80\textwidth,keepaspectratio]{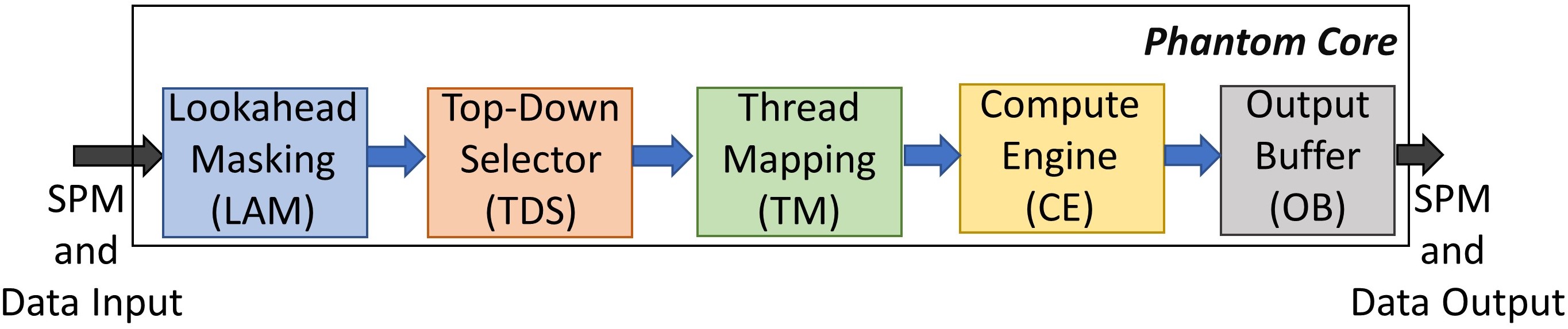}
%%\vspace{-4 mm}
\caption{Phantom block diagram}
\label{arch}
%%\vspace{-3 mm}
\end{figure}

\begin{figure}[ht]
\centering
\includegraphics[width=0.90\textwidth,keepaspectratio]{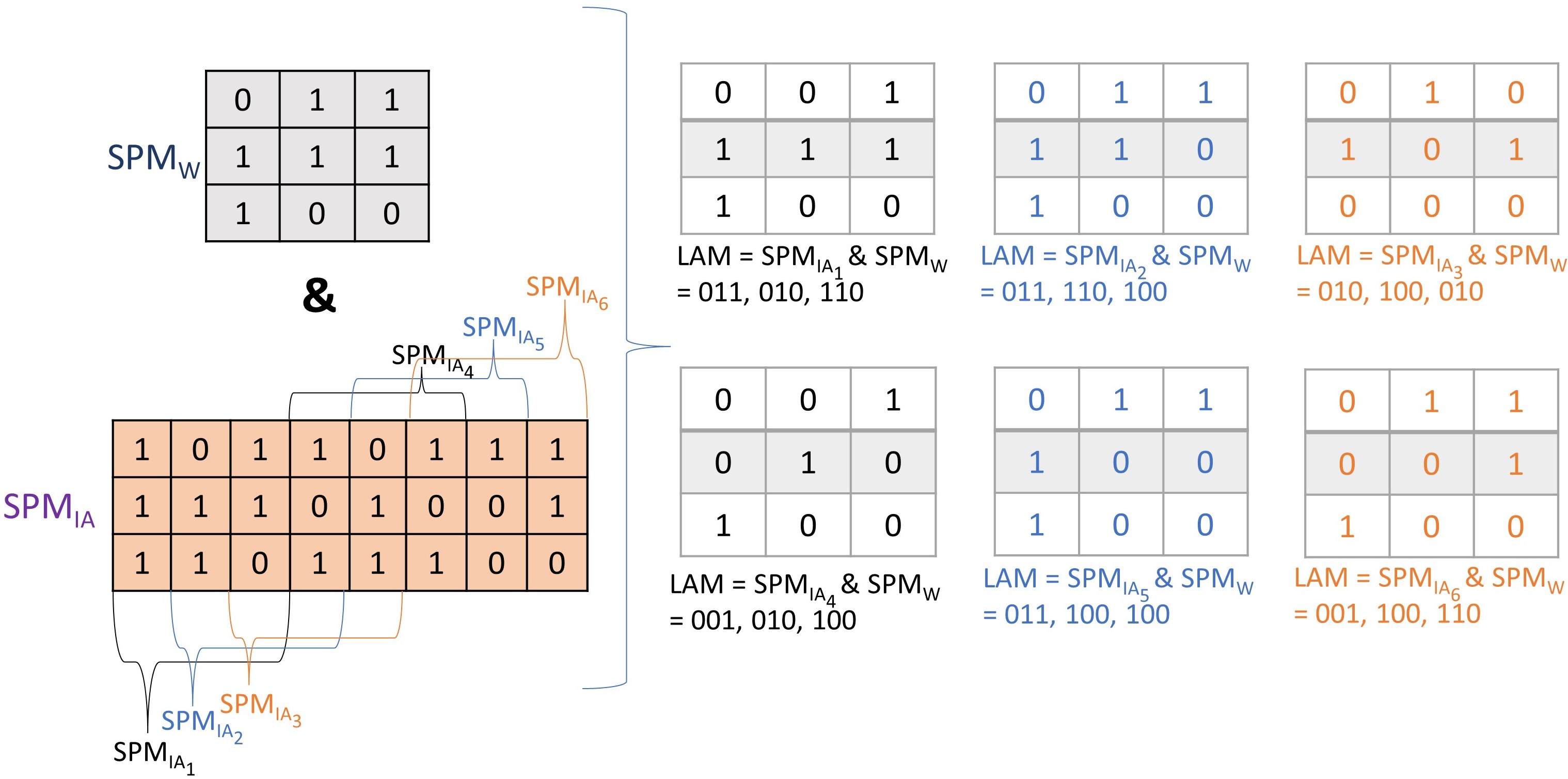}
\caption{ANDing of sparse masks}
\label{anding}
%%\vspace{-3 mm}
\end{figure}
%%%%%%%%%%%%%%%%%%%%%%%%%%%%%%%%%%%%%%%%%%%%%%%%%%%%%%%%%%%%%%%%
%%\vspace{-2 mm}
\subsection{Core Architecture}
%%\vspace{-1 mm}
%%%%%%%%%%%%%%%%%%%%%%%%%%%%%%%%%%%%%%%%%%%%%%%%%%%%%%%%%%%%%%%%
\textit{Phantom} core accepts data in the sparse mask format and consists of 5 main blocks, as shown in Figure \ref{arch}. The \textit{lookahead mask} (LAM) and the \textit{top-down selector} (TDS) blocks use the sparse mask to extract the information about the valid computations and thread selection. The \textit{thread mapper} (TM) uses the information from the previous blocks to efficiently map input and weight data onto the data registers of the multiplier threads within the \textit{compute engine} (CE). The CE consists of an array of \textit{multi-threaded} PEs and level 1 (L1) configurable adders.
%For this design example, we consider a CE array of 3 PEs, with each PE containing 3 threads. This shows that a total of $9$ multiplications can be performed by the CE in parallel. 
The \textit{output buffer} (OB) block consists of an array of FIFO buffers and level 2 (L2) accumulators which generate the final output. 
%Although, the blocks in the core are generic in nature, we will present a specific design example for ease of understanding. The impact of changing design parameters on performance will be explored in section 5.

%%%%%%%%%%%%%%%%%%%%%%%%%%%%%%%%%%%%%%%%%%%%%%%%%%%%%%%%%%%%%%%%

%%%%%%%%%%%%%%%%%%%%%%%%%%%%%%%%%%%%%%%%%%%%%%%%%%%%%%%%%%%%%%%%
%%\vspace{-2 mm}
\subsection{Lookahead Masking}
%%\vspace{-1 mm}
%%%%%%%%%%%%%%%%%%%%%%%%%%%%%%%%%%%%%%%%%%%%%%%%%%%%%%%%%%%%%%%%
Dot product between two vectors is the basic unit of computation in a convolution operation. Considering two-sided sparsity, there are 4 possible multiplication outcomes as a result of sparse vector-vector dot product.\\
A) \textit{zero\textsubscript{w}} $\times$ \textit{zero\textsubscript{a}}\\
B) \textit{zero\textsubscript{w}} $\times$ \textit{non-zero\textsubscript{a}}\\
C) \textit{non-zero\textsubscript{w}} $\times$ \textit{zero\textsubscript{a}}\\
D) \textit{non-zero\textsubscript{w}} $\times$ \textit{non-zero\textsubscript{a}}\\ 
The only valid multiplication is when a non-zero weight data is multiplied by a non-zero activation data. As the name implies, the lookahead mask (LAM) block \textit{looks ahead} into \textit{n} convolution chunks to determine the locations of valid multiplications. We refer to the value $n$ as the \textit{lookahead factor} ($L_f$) \footnote{Notations $n$ and $L_f$ will be used interchangeably throughout the text}. To perform its task, the LAM block performs an AND operation between the sparse masks of the weight matrix and the input chunks to generate output masks. Figure \ref{anding} shows the process of ANDing. Here, SPM\textsubscript{W} and SPM\textsubscript{IA} are the sparse masks of the weight and the input matrix of the example in Figure \ref{chunks}. Six output chunks are generated based on the AND operation.\par
\begin{figure}[t]
\centering
\includegraphics[width=0.70\textwidth,keepaspectratio]{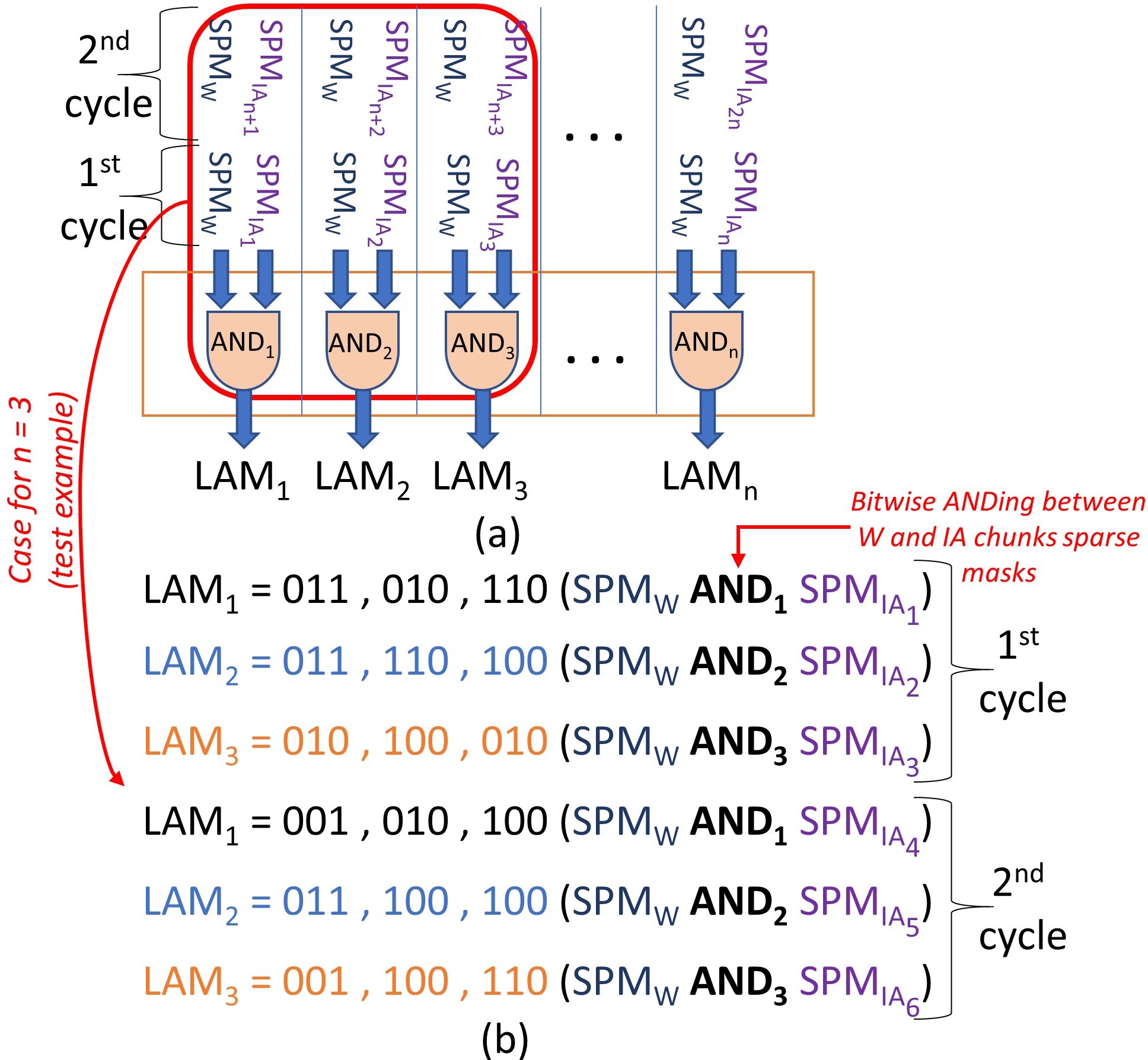}
%%\vspace{-4 mm}
\caption{Lookahead masking (a) Sequence of parallel AND gates (b) Output for test case}
\label{LAM}
%%\vspace{-6 mm}
\end{figure}

To perform this ANDing, the LAM consists of a series of \textit{n} AND gates, as shown in Figure \ref{LAM}(a). Bitwise ANDing is perfomed between the weight and the activation sparse mask. For this example, we set the value of $n$ to 3. This means that to generate six output masks in Figure \ref{anding}, two cycles will be needed for the six AND operations.

Figure \ref{LAM}(b) shows the output of LAM for the test example. The $n=3$ AND gates produce outputs (LAM\textsubscript{1}, LAM\textsubscript{2}, LAM\textsubscript{3}) on every cycle edge. It takes 2 cycles to slide over the entire $3 \times 8$ activation matrix. The ones in the outputs of AND gates represent the location of a valid vector-vector dot product whereas the zeros represent a product resulting in a zero. Overall, it can be seen that by using a sequence of n-AND gates, we can accurately determine the positions of \textit{valid} computations in $n$ convolution chunks. 
%%%%%%%%%%%%%%%%%%%%%%%%%%%%%%%%%%%%%%%%%%%%%%%%%%%%%%%%%%%%%%%%

%%%%%%%%%%%%%%%%%%%%%%%%%%%%%%%%%%%%%%%%%%%%%%%%%%%%%%%%%%%%%%%%
%%\vspace{-1.5 mm}
\subsection{Top-Down Selector}
%%\vspace{-1 mm}
%%%%%%%%%%%%%%%%%%%%%%%%%%%%%%%%%%%%%%%%%%%%%%%%%%%%%%%%%%%%%%%%
The top-down selector (TDS) receives the LAM outputs on every cycle edge and selects a sub-sequence of LAM outputs that can maximize the utilization of multiplier threads in the PEs within the CE. There are a total of \textit{p} parallel selectors, with \textit{p} equal to the total number of PEs in the CE. In this design, we consider $p=3$, since there are a total of 3 PEs in the CE. We develop two selection algorithms for the TDS block namely, \textit{in-order selection} and \textit{out-of-order selection}.

%\begin{figure}[ht]
%\centering
%\includegraphics[width=0.40\textwidth,keepaspectratio]{TDS_2.jpg}
%%%\vspace{-2 mm}
%\caption{In-Order Selection Implementation}
%\label{TDS_2}
%%%\vspace{-2 mm}
%\end{figure}

%%%%%%%%%%%%%%%%%%%%%%%%%%%%%%%%%
%%\vspace{-2 mm}
\subsubsection{In-Order Selection}
\begin{figure}[t]
\centering
\includegraphics[width=0.75\textwidth,keepaspectratio]{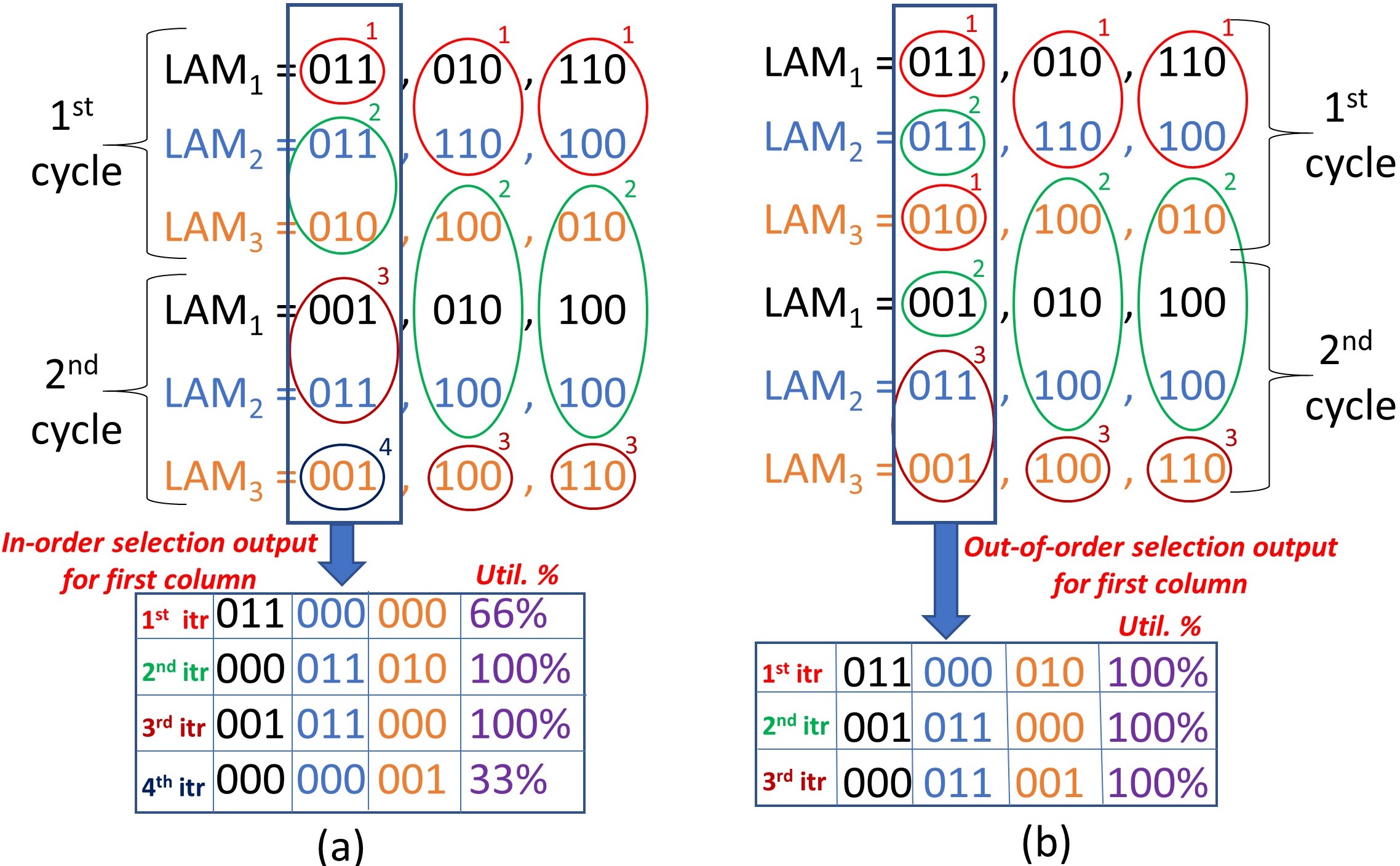}
%%\vspace{-4 mm}
\caption{Top down selector (a) In-order selection (b) Out-of-order selection}
\label{TDS_1}
%%\vspace{-6 mm}
\end{figure}

%%\vspace{-1 mm}
Figure \ref{TDS_1}(a) shows the process of in-order selection. The three selectors work on the three \textit{comma-separated} columns in parallel. The selection is performed iteratively in an \textit{ordered} fashion on the entire column in a top-down manner. For column 1 in Figure \ref{TDS_1}(a), the first selector loops through the first $n$ elements in the current iteration, i.e., $011$(black), $011$(blue), and $010$(orange). Here, $n$ equals the \textit{lookahead factor} ($L_f$), which is 3 in this design. The first $011$ (in black) is assigned the highest priority and is selected. The selector counts the number of ones from this entry and stores the result. The selector then proceeds to the next entry, i.e., $011$ (in blue), and counts the number of ones. If the combined sum of the number of ones of the current and the previous entry is greater than $n$, the current \textbf{and} the next entries, in the current iteration are not considered, and instead, replaced by zeros. Here, we see that 011(in black) + 011(in blue) = 4 > ($n=3$). Therefore, only the first 011 (in black) is selected and the rest two entries are replaced by zeros in the first iteration. Here, every 1 in the selected output corresponds to a valid multiplication. Since, there are a total of 3 multiplier threads within a single PE,
if the selected values contain a total of three 1s, then all the threads within a PE are utilized, giving a utilization of 100\% (as shown in Figure \ref{TDS_1}(a)). The utilization decreases if the order in which the entries appear do not align with the priority of selection. The circles around the values and the numbers on top represent the iteration number during which the particular value was selected by the selector. In the second iteration, the selector goes on to the first unselected value (011 in blue) and follows the same selection process. It takes a total of four iterations (cycles) for the selector to select all the values in the first column, as shown in Figure \ref{TDS_1}(a). We can also see that the second and the third columns require a total of three cycles for selection, but need to wait one additional cycle for the selection of the first column to complete. This can cause computational idling and under-utilization of the PE multipliers.
%
%%\vspace{-3 mm}

\subsubsection{Out-of-Order Selection}
\begin{figure}[t]
\centering
\includegraphics[width=0.80\textwidth,keepaspectratio]{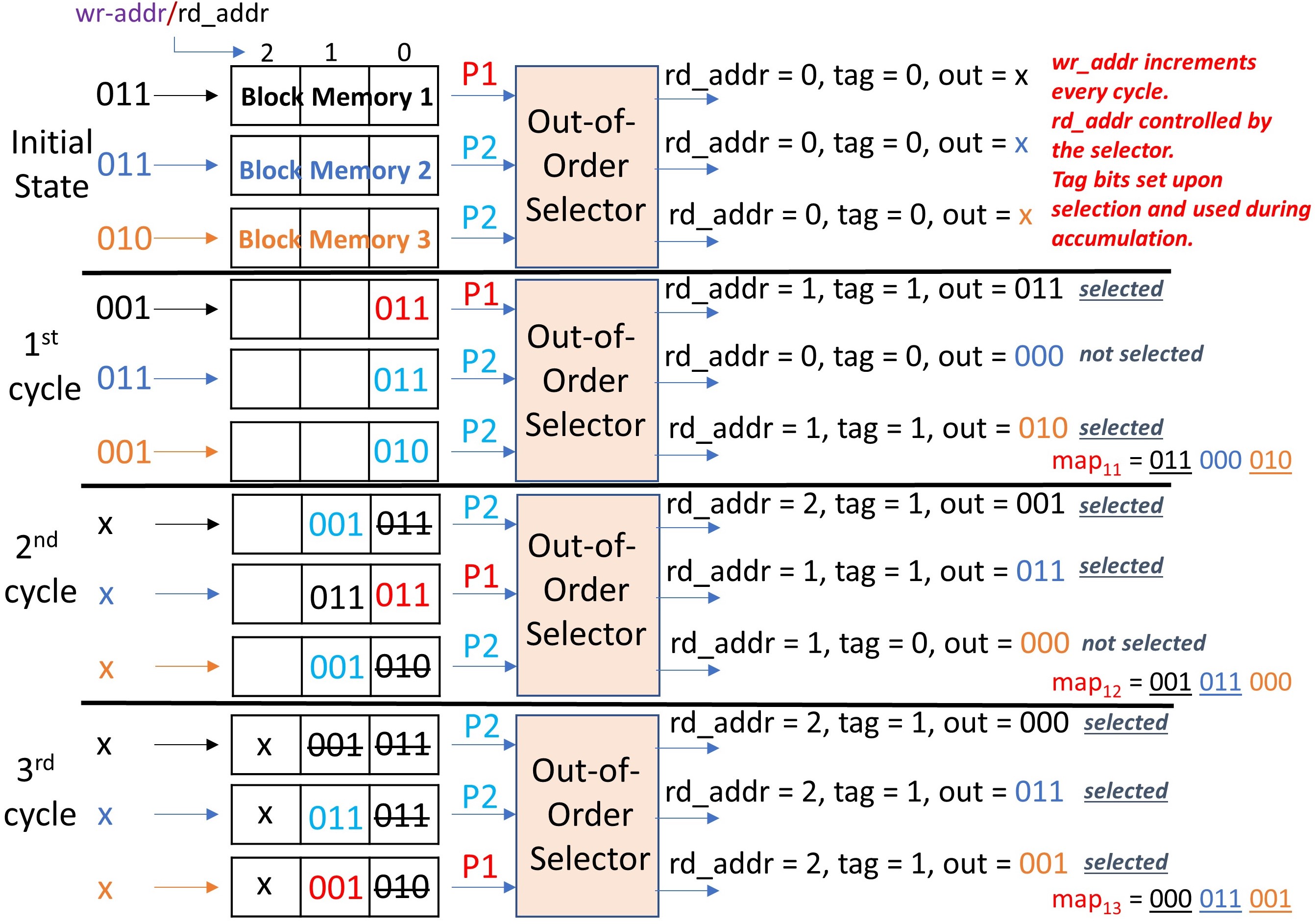}
%%\vspace{-2 mm}
\caption{Out-of-order selection implementation}
\label{TDS_3}
%%\vspace{-3 mm}
\end{figure}

\begin{figure}[t]
\centering
\includegraphics[width=0.70\textwidth,keepaspectratio]{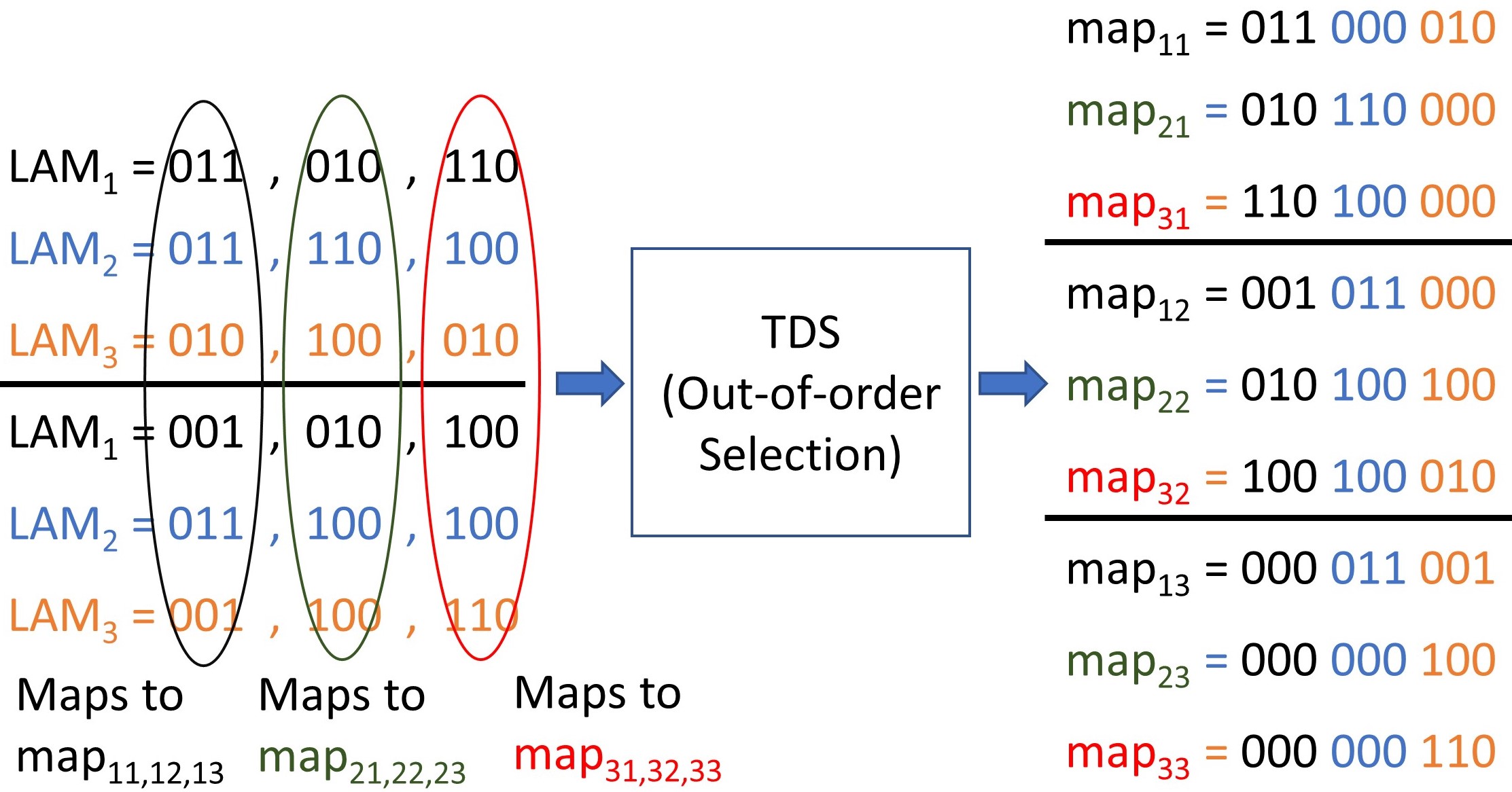}
%%\vspace{-2.5 mm}
\caption{TDS out-of-order variant input-output}
\label{TDS_4}
%%\vspace{-6.0 mm}
\end{figure}

%%\vspace{-1 mm}
In-order selection is highly dependent on the order in which the inputs appear. This can lead to under-utilization of the multiplier threads in the PE, as shown in Figure \ref{TDS_1}(a). Figure \ref{TDS_1}(b) shows the out-of-order selection method, where after the selection of the 1\textsuperscript{st} entry (011 in black), \textbf{all} next entries (011 in blue, 010 in orange) in an iteration are considered for selection. Therefore, as shown in Figure \ref{TDS_1}(b), in the first iteration, after the selection of 011(in black), the next value 011 (in blue) is not considered (011 + 011 = 4 > $n$) but the subsequent value 010 (in orange) is considered because 011(in black) + 010 (in orange) $\leq n$ . This small, yet efficient change greatly improves the thread utilization and consequently the throughput, as shown in Figure \ref{TDS_1}(b). Figure \ref{TDS_3} shows the implementation of the out-of-order selection variant of the TDS for the first column in Figure \ref{TDS_1}(b).
A small block memory, having independent read and write ports, receives the LAM outputs on every cycle. P1 and P2 are the high and low priorities, respectively. The read address (rd\textunderscore addr) of the memory increments every time a value is selected and the tag bits are set to 1 for those values. The tag bits serve the purpose of accumulation during the output buffering (explained in Section~\ref{SectionOB}). The priority of selection is reversed on the next input to ensure that the values missed in the previous iteration are given the highest priority in the current iteration. \textit{map\textsubscript{11}}, \textit{map\textsubscript{12}}, and \textit{map\textsubscript{13}} are the final outputs, %(select\textunderscore matrix in Algorithm \ref{inOrder} and Figure \ref{TDS_1}(a))
as shown in Figure \ref{TDS_4}. 
%After exhausting the entire column, all the rd\textunderscore addr's are equal and the tag bits set to 1. 
%In this case, there are only two priorities so both the entries after the  highest priority are equally considered. This ensures that the thread utilization is maximized. 
The hardware overhead of the out-of-order selector is roughly 1.03$\times$ that of the in-order selector variant, but can increase the hardware utilization by as much as 60\%. \textit{Phantom}, therefore, employs the out-of-order selection variant of the TDS for higher thread utilization and efficient mapping.
%%%\vspace{-3 mm}
%
%%%\vspace{-2 mm}
\begin{figure*}
\centering
\includegraphics[width=0.99\textwidth,keepaspectratio]{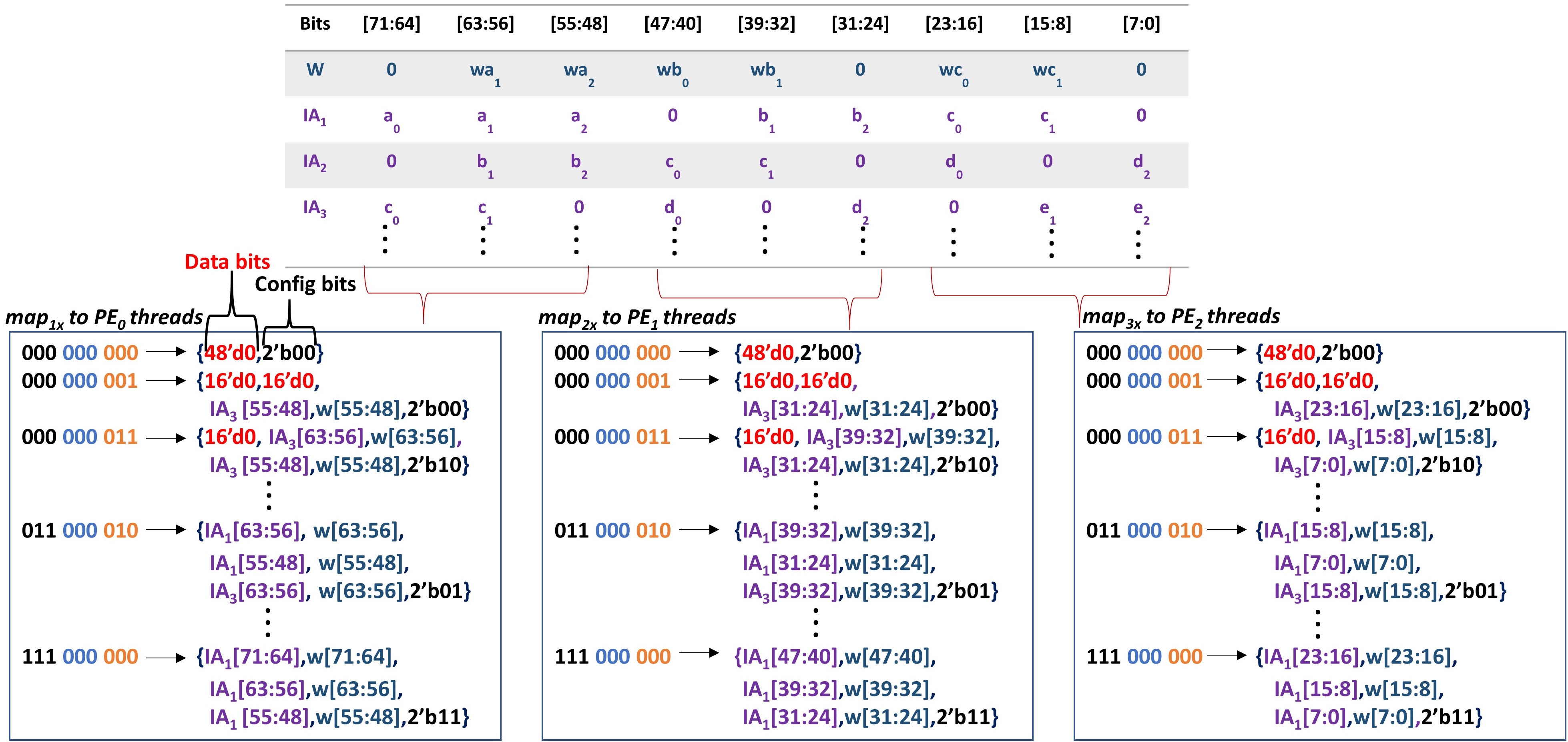}
%%\vspace{-4 mm}
\caption{Thread mapping}
\label{map}
%%\vspace{-4 mm}
\end{figure*}
%%%\vspace{-3 mm}
%

%%%%%%%%%%%%%%%%%%%%%%%%%%%%%%%%%%%%%%%%%%%%%%%%%%%%%%%%%%%%%%%%

%%%%%%%%%%%%%%%%%%%%%%%%%%%%%%%%%%%%%%%%%%%%%%%%%%%%%%%%%%%%%%%%
%%\vspace{-2 mm}
\subsection{Thread Mapper}
%%\vspace{-1 mm}
%%%%%%%%%%%%%%%%%%%%%%%%%%%%%%%%%%%%%%%%%%%%%%%%%%%%%%%%%%%%%%%%
The Thread Mapper (TM) takes the input from the TDS (\textit{map\textsubscript{1x}, \textit{map\textsubscript{2x}}, and \textit{map\textsubscript{3x}}}) and uses this information to map the length equalized sparse data (weight and activation) onto the internal registers of the multi-threaded PEs. The length equalization is done by adding appropriate zeros at specific locations using the sparse masks. Figure \ref{map} shows the map operation for the three PEs in this design. 
%Even though there are a total of
Out of the 2\textsuperscript{9} = 512 combinations, the mapper only needs to store those for which the total number of ones do not exceed the multiplier count of each PE (3 in this case). This drastically reduces the total combinations needed to be stored (${9 \choose 0} + {9 \choose 1} + {9 \choose 2} + {9 \choose 3} $ = 130, a 74\% reduction in the memory footprint). Each PE has a 50 bit internal register, out of which, 48 bits are the data bits, and 2 bits are the L1 adder control bits. The first 48 bits are divided into a set of 16 bits (8 bits for both activations and weights). The total memory requirement for storage of all three mappers is approximately 2.5 kB. One key observation from Figure \ref{map} is that the mapper 2 and 3, map the data in a similar fashion as the mapper 1, but only use different location bits for weights and data. We, therefore, remove the two mappers (map\textsubscript{2x} and map\textsubscript{3x}), and only use one mapper (map\textsubscript{1x}) sequentially 3 times, and adjusting the location bits afterwards. This only incurs an initial latency of 2 cycles but reduces the memory footprint by approximately 66\% (2.5 kB to 0.83 kB). Appropriate delay registers are added in the PEs to account for the initial delay.    

%%%%%%%%%%%%%%%%%%%%%%%%%%%%%%%%%%%%%%%%%%%%%%%%%%%%%%%%%%%%%%%%

\begin{figure}[ht]
\centering
\includegraphics[width=0.70\textwidth,keepaspectratio]{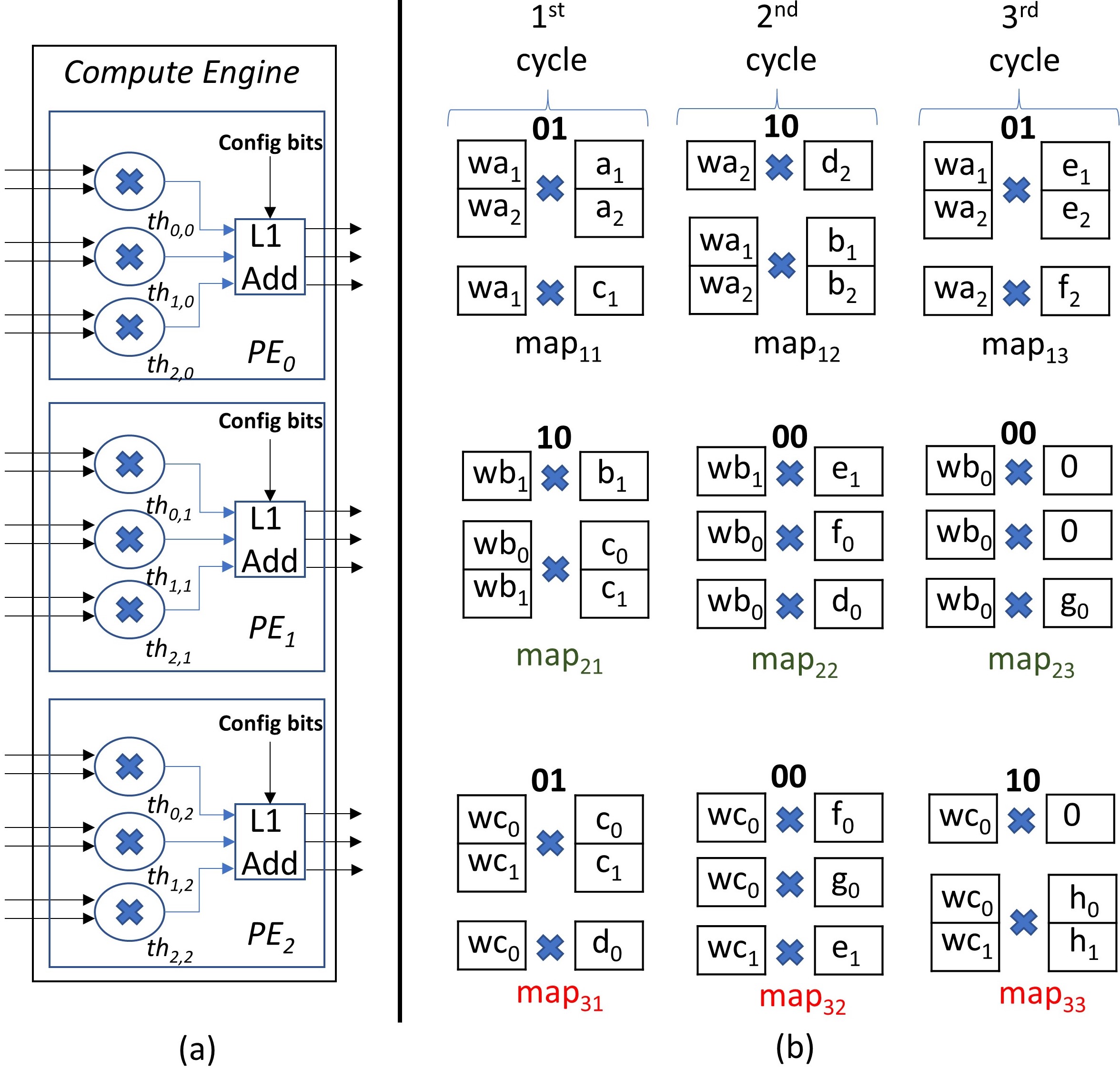}
%%\vspace{-4.9 mm}
\caption{Compute engine (a) Multi-threaded design (b) Scheduling for example in Figure \ref{chunks}}
\label{CE}
%%\vspace{-6.5 mm}
\end{figure}
%%%\vspace{-2 mm}
%%%%%%%%%%%%%%%%%%%%%%%%%%%%%%%%%%%%%%%%%%%%%%%%%%%%%%%%%%%%%%%%
%%\vspace{-2 mm}
\subsection{Compute Engine}
%%\vspace{-1 mm}
%%%%%%%%%%%%%%%%%%%%%%%%%%%%%%%%%%%%%%%%%%%%%%%%%%%%%%%%%%%%%%%%
The \textit{Phantom} core uses a multi-threaded CE block. In this particular design, the CE block consists of 3 multi-threaded PEs, with each PE containing 3 multiplier threads, as shown in Figure \ref{CE}(a). The mapper maps the data to the individual threads which perform independent computations. The outputs of the threads, local to a particular PE, are provided to the L1 adder. The L1 adder is provided the configuration bits from the last 2 bits of the mapper (Figure \ref{map}). There are 4 cases for the configuration bits:\\
\textit{C1: 00 -> The individual outputs of the multiplier threads within the PE are not added and simply passed.}\\
\textit{C2: 01 -> The outputs of the first two multiplier threads (th\textsubscript{0,x},th\textsubscript{1,x}) are added, whereas, the third one (th\textsubscript{2,x}) is passed as is.  }\\
\textit{C3: 10 -> the outputs of th\textsubscript{1,x} and th\textsubscript{2,x} are added and the output of th\textsubscript{0,x} is passed as is.  }\\
\textit{C4: 11 -> the outputs of all the multiplier threads are added.  }\\
Figure \ref{CE}(b) shows the cycle by cycle scheduling of the multiplier threads with data (weights and activations) mapped using the logic in Figure \ref{map}. The mapping inputs (map\textsubscript{11}, etc.) and L1 adder configuration outputs are also shown. It can be seen that all the multiplier threads are efficiently utilized even though the output is 55\% sparse (counting number of zeros from the left side of Figure \ref{TDS_4}, 30/54). The hardware utilization during the 1\textsuperscript{st} and 2\textsuperscript{nd} cycle is 100\%, whereas, for the 3\textsuperscript{rd} cycle, it is 66\%. The reason for low utilization in the last cycle is because the input is at the \textit{boundary} and the LAM block does not have more data to \textit{look ahead} into.
%%%%%%%%%%%%%%%%%%%%%%%%%%%%%%%%%%%%%%%%%%%%%%%%%%%%%%%%%%%%%%%%

\begin{figure}[ht]
\centering
\includegraphics[width=0.70\textwidth,keepaspectratio]{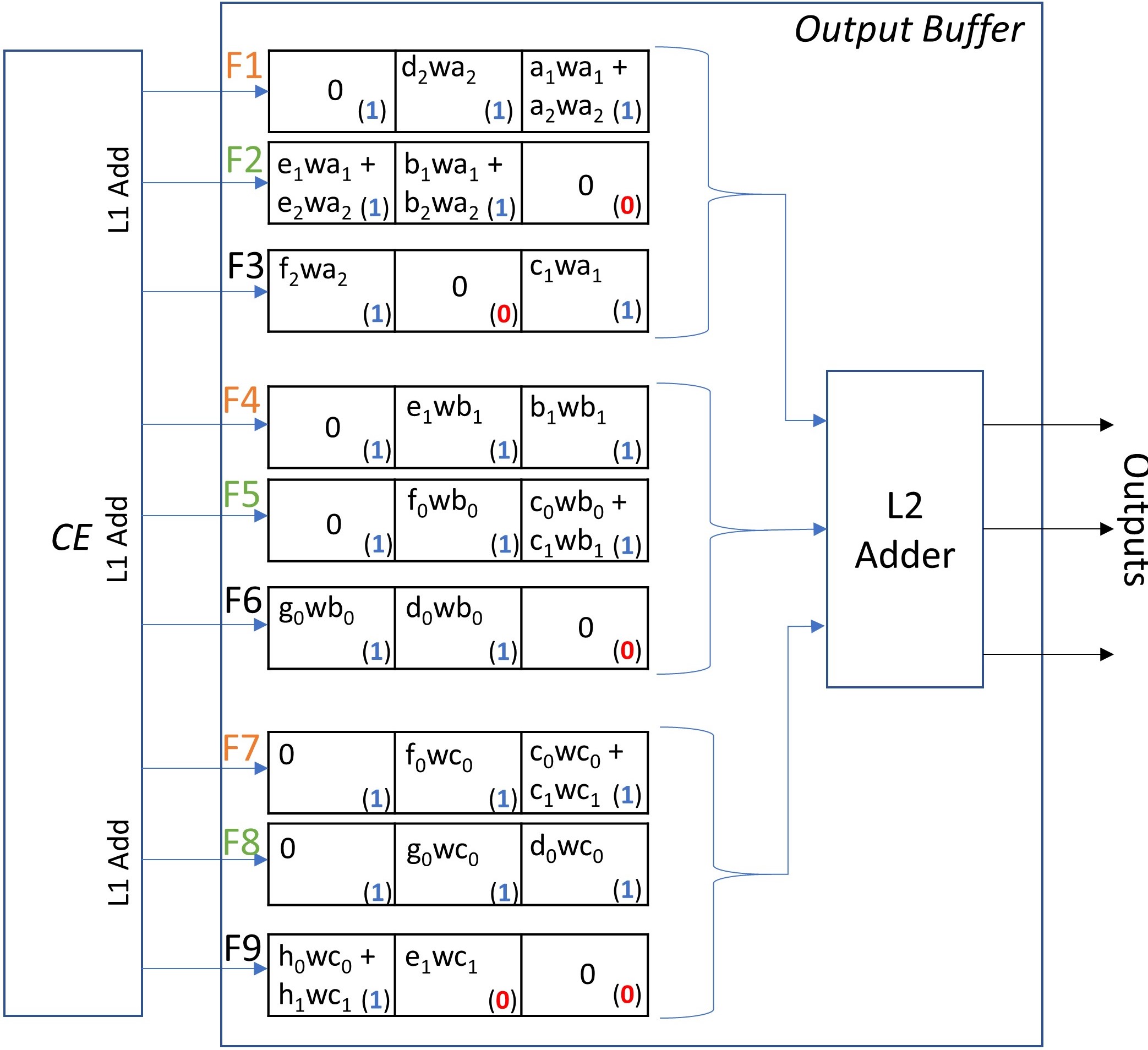}
%%\vspace{-3 mm}
\caption{Output buffer for example in Figure \ref{chunks}}
\label{OB}
%%\vspace{-6 mm}
\end{figure}

\begin{figure}[t]
\centering
\includegraphics[width=0.60\textwidth,keepaspectratio]{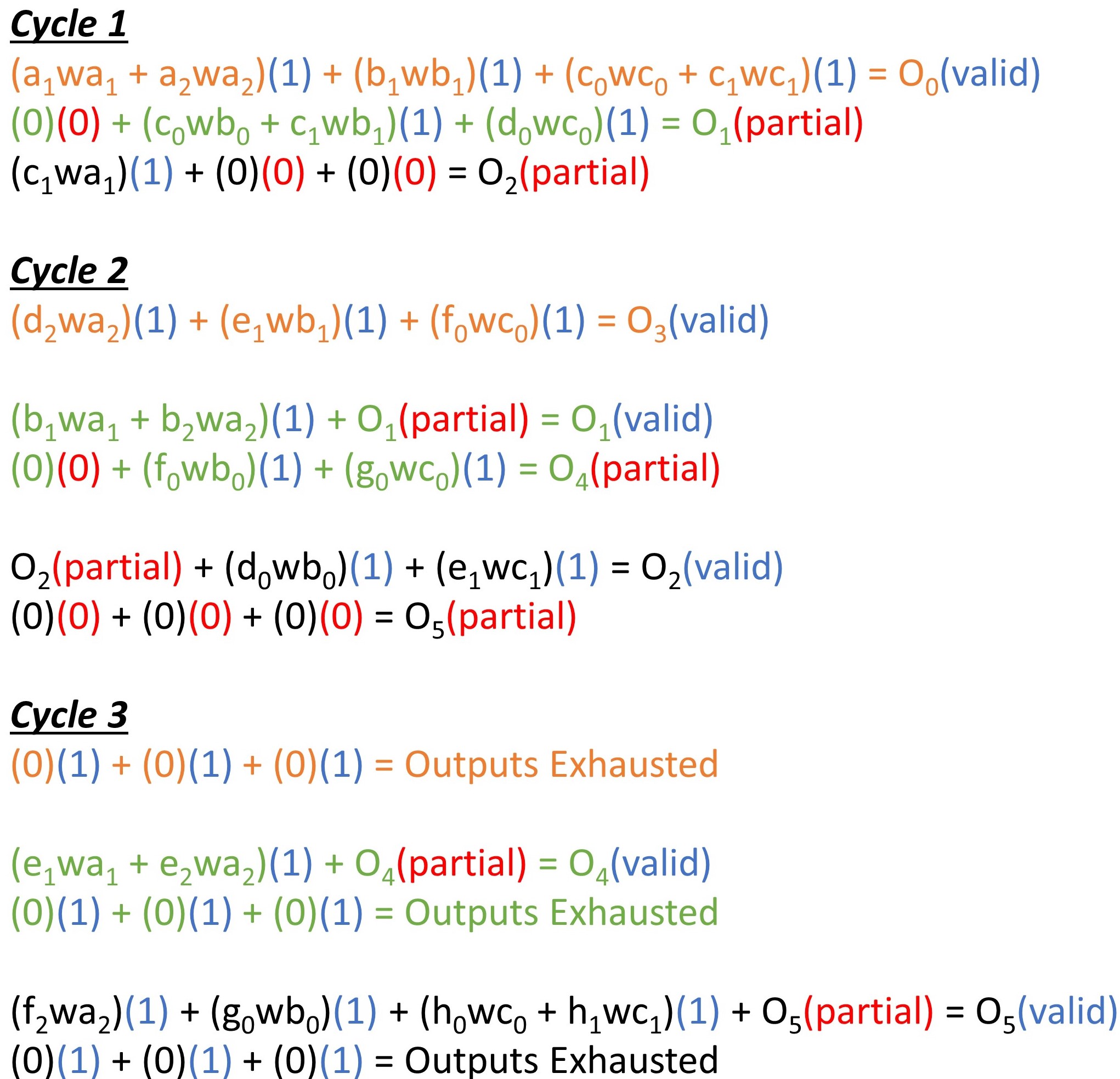}
%%\vspace{-3 mm}
\caption{L2 accumulation for the example in Figure \ref{chunks}}
\label{OB_2}
\vspace{5 mm}
\end{figure}
%%%%%%%%%%%%%%%%%%%%%%%%%%%%%%%%%%%%%%%%%%%%%%%%%%%%%%%%%%%%%%%%
%%\vspace{-2 mm}
\subsection{Output Buffer} \label{SectionOB}
%%\vspace{-1 mm}
%%%%%%%%%%%%%%%%%%%%%%%%%%%%%%%%%%%%%%%%%%%%%%%%%%%%%%%%%%%%%%%%
The final block in the \textit{Phantom} core is the output buffer (OB). OB is responsible for buffering the outputs of the CE, and accumulation of data using the L2 adder to generate the final outputs. The buffering is performed using a system of $m$ first-in, first out (FIFO) buffers, where $m$ is equal to the total number of multiplier threads in the CE (9). Figure \ref{OB} shows the OB block for the example in Figure \ref{chunks}. F1-F9 are the 9 fifos receiving data from the L1 adders in the CE. The \textit{ones} (in blue) and \textit{zeros} (in red) within the parenthesis represent the \textit{tag} bits which were set by the TDS block, as shown in Figure \ref{TDS_3}. Accumulation is performed in two stages by the same colored fifos (F1 + F4 + F7), (F2 + F5 + F8), and (F3 + F6 + F9). The outputs are either \textit{valid} or \textit{partial}, based on the associated tags. If the tags of all the values being accumulated are equal to \textbf{1}, the output is considered \textit{valid}, otherwise, it is considered \textit{partial}. For the example in Figure \ref{chunks}, the outputs in L2 adder are calculated as shown in Figure \ref{OB_2}. The two stages of accumulation are also shown. In the first stage, the previously accumulated \textit{partial} output is added to the new entry in the fifo to make the output \textit{valid}. This is done by checking the tag bits in the \textit{partial} output and adding the missing  tag \textbf{1} in the new value to generate the \textit{valid} output from \textit{partial} output. In the second stage, new partial values are generated by replacing the used tag \textbf{1} value in the first stage by 0 and accumulating the rest of the tag \textbf{1} values. The process ends once the input is exhausted and all the tag \textbf{1} values have been accumulated to generate \textit{valid} outputs. \par

\begin{figure}
\centering
\includegraphics[width=0.65\textwidth,keepaspectratio]{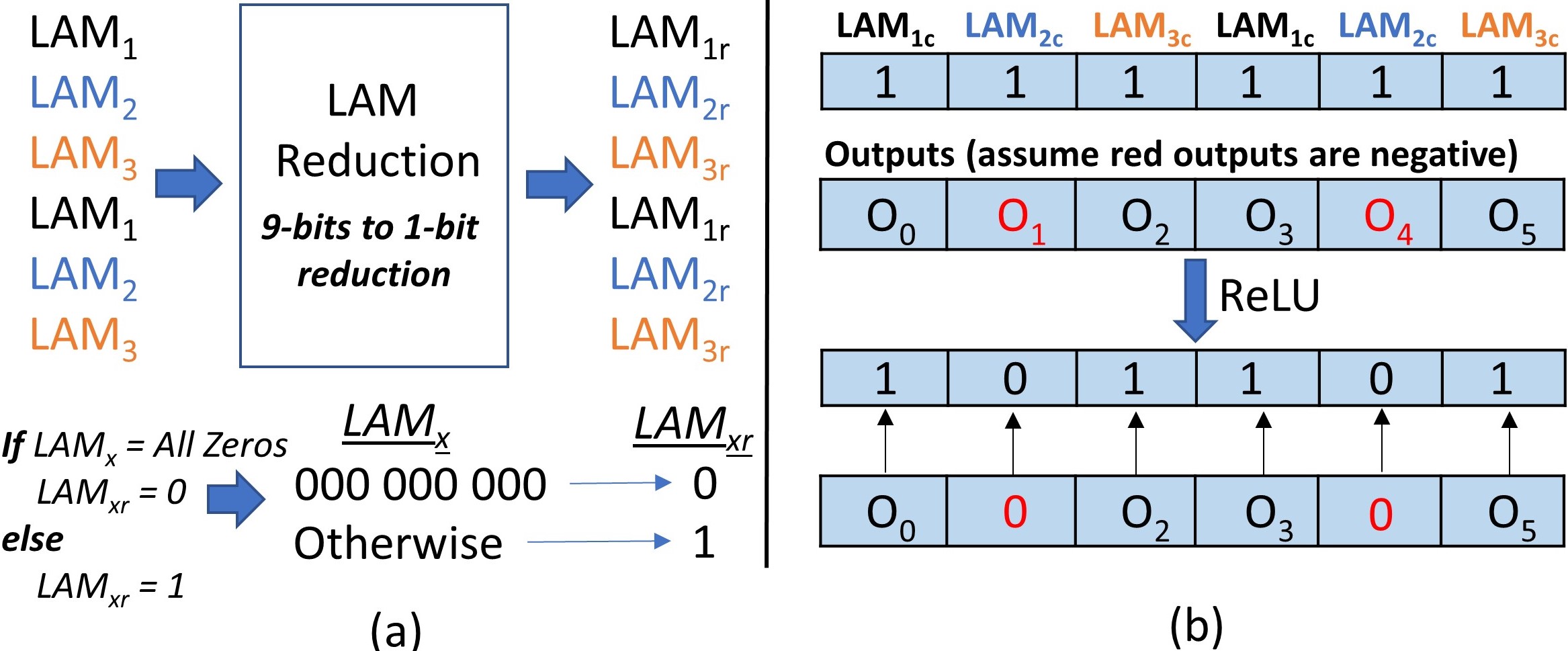}
%\vspace{-4 mm}
\caption{Output sparse mask generation}
\label{encode}
%\vspace{-7 mm}
\end{figure}

%%%%%%%%%%%%%%%%%%%%%%%%%%%%%%%%%%%%%%%%%%%%%%%%%%%%%%%%%%%%%%%%%%%%%%%%%
\subsection{Output Encoding}
%%%%%%%%%%%%%%%%%%%%%%%%%%%%%%%%%%%%%%%%%%%%%%%%%%%%%%%%%%%%%%%%%%%%%%%%%
Figure~\ref{encode} shows the process of output sparse mask generation. Unlike the weights, the output activation sparsity is dynamic and the sparse mask needs to be generated on-the-fly. From Figure~\ref{anding}, we can see that the presence of even a single \textit{one} in the LAM outputs represent a non-zero output. To determine the output sparse mask, the same metadata can be used. The first step involves reduction of the individual LAM outputs to a single bit (LAM\textsubscript{xr}), based on an \textit{all-zero} check, as shown in Figure~\ref{encode}(a). This generates the sparse mask for the outputs before ReLU. Note that the LAM values are taken from the test example (Figure~\ref{chunks}). Figure~\ref{encode}(b) shows the second step after ReLU, where the negative outputs, and their corresponding sparse mask locations, are converted into zeros. This final sparse mask is stored as is, whereas, the output is shifted first to omit zero data entries, and then stored.\par

This concludes the processing in a single \textit{Phantom} core. In the next section, we will introduce \textit{Phantom-2D}; a two-dimensional accelerator having a system of \textit{Phantom} cores for processing CNN layers during the inference process.

%%%%%%%%%%%%%%%%%%%%%%%%%%%%%%%%%%%%%%%%%%%%%%%%%%%%%%%%%%%%%%%%
\begin{figure}[ht]
\centering
\includegraphics[width=0.70\textwidth,keepaspectratio]{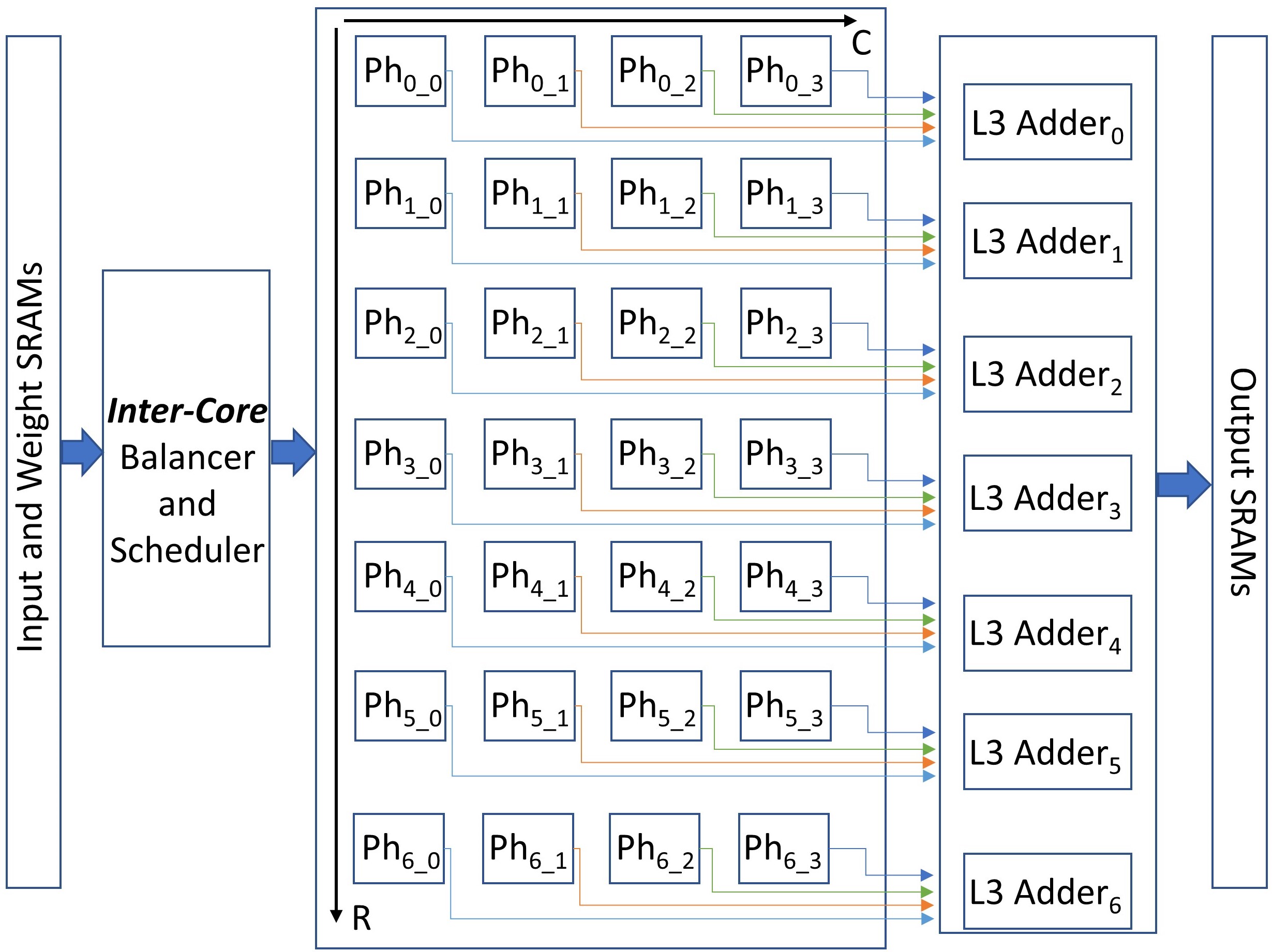}
%%\vspace{-2 mm}
\caption{Phantom-2D architecture}
\label{ph3d}
%%\vspace{-6 mm}
\end{figure}

%%%%%%%%%%%%%%%%%%%%%%%%%%%%%%%%%%%%%%%%%%%%%%%%%%%%%%%%%%%%%%%%
%%\vspace{-3 mm}
\section{Phantom-2D}
%%\vspace{-1 mm}
%%%%%%%%%%%%%%%%%%%%%%%%%%%%%%%%%%%%%%%%%%%%%%%%%%%%%%%%%%%%%%%%
The previous sections describe the working of the various blocks in the \textit{Phantom} core. The individual core works by computing a dot product between a weight matrix and a subsection of the input feature map to compute a subsection of the output feature map. To compute the \textit{full} output feature map, corresponding to a particular layer, we design a two-dimensional accelerator comprising of the \textit{Phantom} cores, which we refer to as \textit{Phantom-2D}, as shown in Figure \ref{ph3d}. We envision that the \textit{Phantom-2D} accelerator connects to a memory bus and accepts data and instructions from a CPU. The accelerator contains on-chip SRAMs (weight, input, and output), the \textit{inter-Core} balancer and scheduler, the $R\times C$ compute matrix, comprising of \textit{Phantom} cores, and the accumulator circuits. 
%Based on the input layer, \textit{Phantom-2D} can perform CONV operations (regular and separable) as well as FC layer operations to support the complete structure of modern CNNs.

%\begin{scriptsize}
%\begin{table}[ht]
%  \centering
%  \begin{tabular}{|l|l|l|}
%    \hline
%    \textbf{CNN} & \textbf{Layer Dims} & \textbf{Channel count}\\
%    \hline
%    \hline
%    \multirow{6}{*}{VGG-16} &  $3\times [224\times224]$ & $[3]$\\
%    & $2\times [112\times112]$ & $2\times [64]$ \\
%    & $3\times [56\times56]$ & $2\times [128]$\\
%    & $3\times [28\times28]$ & $3\times [256]$\\
%    & $3\times [14\times14]$ & $3\times [512]$ \\
%    & $4\times [7\times7]$ & $3\times [512]$\\
%    \hline
%    \multirow{7}{*}{MobileNet} &  $1\times [224\times224]$ & $[3]$\\
%    & $3\times [112\times112]$ & $2\times [32]$ \\
%    & $4\times [56\times56]$ & $2\times [64]$\\
%    & $4\times [28\times28]$ & $4\times [128]$\\
%    & $4\times [14\times14]$ & $4\times [256]$ \\
%    & $4\times [7\times7]$ & $4\times [512]$\\
%    &  & $2\times [1024]$\\
%    \hline
%  \end{tabular}
%  \caption{CNN Layer Dimensions }
%  \label{layers}
%\end{table}
%\end{scriptsize}

%%%%%%%%%%%%%%%%%%%%%%%%%%%%%%%%%%%%%%%%%%%%%%%%%%%%%%%%%%%%%%%%
%%\vspace{-1 mm}
\subsection{R$\times$C Compute Matrix}
%%\vspace{-1 mm}
%%%%%%%%%%%%%%%%%%%%%%%%%%%%%%%%%%%%%%%%%%%%%%%%%%%%%%%%%%%%%%%%
The compute unit consists of an $R\times C$ matrix of the \textit{Phantom} cores and $R$ adders for channel accumulation.
%%%\vspace{-2 mm}
%\subsubsection{$R\times C$ Matrix}
The design choice for $R$ and $C$, and the dataflow associated with the transfer of data across various \textit{Phantom} cores, is based on the following design goals:
\begin{itemize}
    \item [G1:] To maximize the data reuse (weights and input) across multiple input subsections.
    \item [G2:] To optimize the data scheduling across various \textit{Phantom} cores to maximize the \textit{theoretical} hardware utilization.
    \item [G3:] To support \textbf{all} layers of a CNN. This includes support for a variety of CONV layers and the FC layers.
\end{itemize}

Looking deeper into the dimensions and configurations of various popular CNN models (VGGnet \cite{vgg16}, Resnet\cite{resnet}, MobileNets\cite{mobnetv1,mobnetv2}, Googlenet\cite{googlenet} etc), we observe that the channel count for various CNN layers is almost always a multiple of 4. In the \textit{Phantom-2D} architecture, the channels in a CNN layer are broken along the columns. Therefore, to ensure that the cores are engaged \textit{most} of the time, and thus, satisfy G2, we set the $C$ value equal to $4$. Similar rationale applies for the choice of $R=7$. The input subsections are broken along the vertical axis and distributed along the rows. Most of the popular CNNs have an input layer size (width or height) $S$, a multiple of $7$, thus, having $R=7$, ensures equal distribution of chunks of data among all the cores. 

%\begin{figure*}
%\centering
%\includegraphics[width=0.73\textwidth,keepaspectratio]{fcn.jpg}
%%\vspace{-2 mm}
%\caption{Support for different layers with examples (a) Regular/Depthwise Convolution (b) 1D (pointwise) Convolution (c) FC Layer }
%\label{fcn}
%%\vspace{-5 mm}
%\end{figure*}

%%\vspace{-2 mm}
\subsection{Load Balancing}
%%\vspace{-1 mm}
The design choice for $R\times C$ matrix is highly dependent on the layer dimensions of various CNNs. It, however, has no relevance to the static and dynamic sparsity in weights and inputs, respectively. 
%Computations are generally imbalanced among the processing elements (PEs) in an NNA that actively exploits sparsity. 
Efficient reuse of data is one of the key requirements to minimize the memory accesses, repeatedly, for the same data. This reuse, however, amplifies the computational imbalance among different PEs. If the same filter is held in the local memory of the PEs and the input subsections are swept across the PEs, the subsections with higher density would inevitably take more cycles to compute the output, compared to the PEs receiving the subsections with lower density. This varying sparsity of the input maps would create a \textit{system-level} load imbalance among the PEs which would be exposed during the next filter broadcast. Holding the input maps and sweeping the filters would also have similar results, as would the buffering of data. In order to address this \textit{system-level} load imbalance, we incorporate a two-level load balancing strategy in the \textit{Phantom} architecture. The \textbf{Inter-Core} balancer (Figure \ref{ph3d}), balances the computational load dynamically using the density of the weight matrix such that each phantom core works on the weight matrices with the same/similar density over the CNN layers (as described in Section~\ref{SectionInterCoreBalancing} with an example). This balancing is only performed when the weight data is actively being reused (e.g. in regular and depthwise seperable CONV). 
%After performing the balancing operation, the balancer schedules the data on to the \textit{Phantom-2D}. 
The second balancer, referred to as \textbf{Intra-Core} balancer, is local to each core, and performs column-wise balancing of the weight matrix. This balancer is always enabled, regardless of the layer, and significantly improves the individual multiplier thread scheduling performed by the TDS.

 %Finally, it should be noted that the choice of $R$ and $C$ is also based on ensuring a moderate hardware resource requirements (LUT count in FPGAs) and the chip area(mm\textsuperscript{2} in ASICs). 
 
 In the subsequent sections, we demonstrate, with examples, the designed dataflow for various CONV and FC layers. We choose the input sizes that fit well with the layer sizes of actual CNNs. We also show how the balancing is performed to prevent idling of the cores, which in turn, maximizes the throughput, all the while ensuring high data reuse.
 %We also show how our design choice of $R\times C$ matrix maximizes the theoretical hardware utilization and provides support for different layers, thereby, satisfying the design goals G1, G2, and G3.
%%%%%%%%%%%%%%%%%%%%%%%%%%%%%%%%%%%%%%%%%%%%%%%%%%%%%%%%%%%%%%%%

\begin{figure}[ht]
\centering
\includegraphics[width=0.65\textwidth,keepaspectratio]{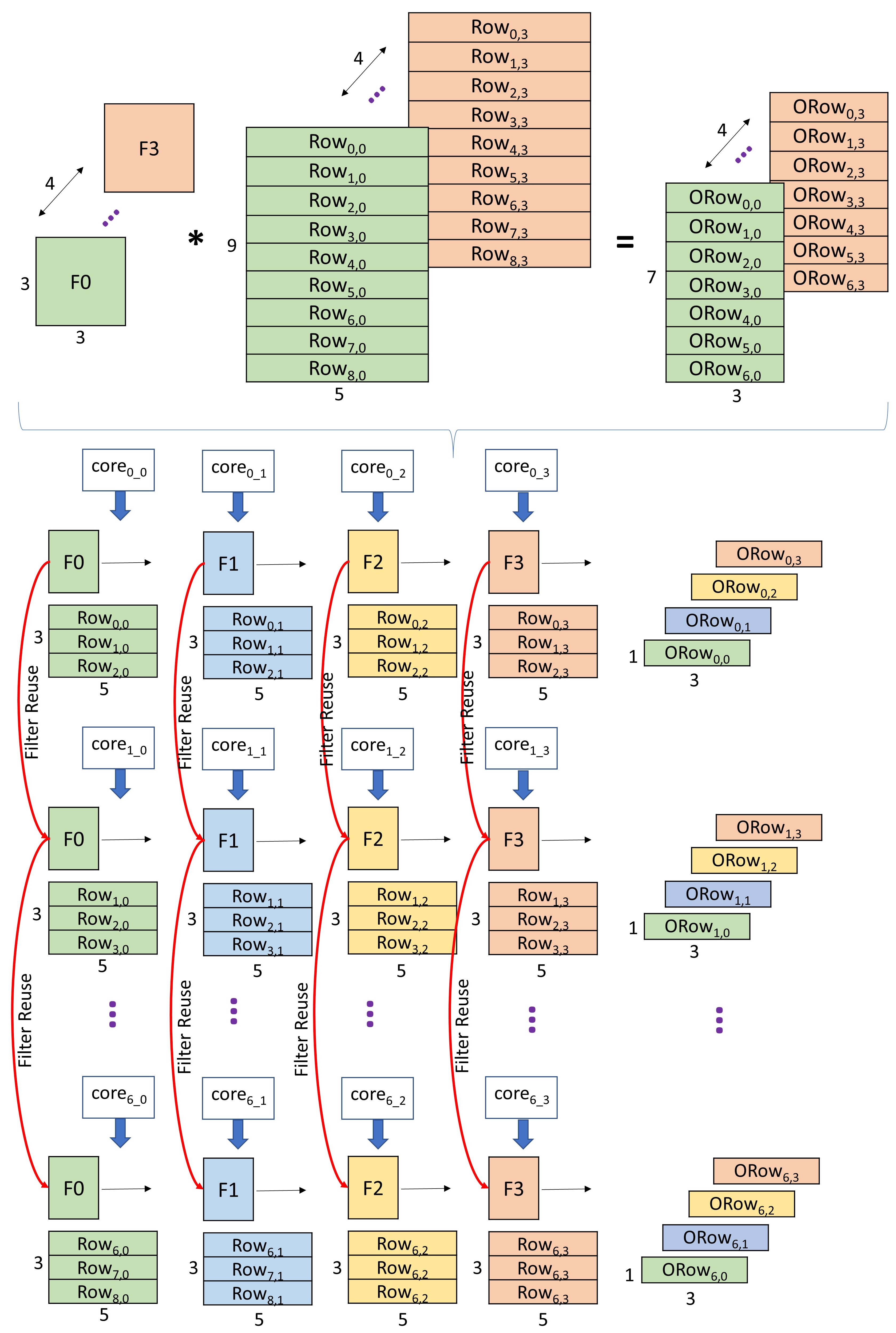}
\caption{Regular/depthwise convolution}
\label{conv1}
\end{figure}

%%%%%%%%%%%%%%%%%%%%%%%%%%%%%%%%%%%%%%%%%%%%%%%%%%%%%%%%%%%%%%%%
%%\vspace{-1 mm}
\subsection{Regular/Depthwise Convolution}
%%\vspace{-1 mm}
%%%%%%%%%%%%%%%%%%%%%%%%%%%%%%%%%%%%%%%%%%%%%%%%%%%%%%%%%%%%%%%%
Figure \ref{conv1} shows a $3\times3$ depthwise separable convolution example, where 4, $3\times3$ filters are convolved with a $9\times5\times4$ input to generate a $7\times3\times4$ output. We choose these size parameters as an example because they fit well when considering the layer sizes of the actual CNNs. The Figure also shows the scheduling and mapping of data on to the $R\times C$ matrix. The input is broken down into $n$ chunks along the rows, where, $n$ is the number of rows of the output. These $n$ chunks are then scheduled along the rows of the $R\times C$ matrix. Each \textit{Phantom} core processes a $3\times5$ input chunk to generate a $1\times3$ output chunk.
%The \textit{Phantom} cores along the columns process channels. 
The filters F0 to F3 represent the channel wise filters, with each column of the $R\times C$ matrix processing a different channel. The reuse of filters (G1) along the rows is also shown. The non-unit stride convolutions follow the same dataflow. Because of the efficient dataflow, and choice of the $R\times C$ dimensions, all \textit{Phantom} cores are provided data for a particular processing chunk, thereby, achieving a 100\% inter-core utilization (G2).\par

%%\vspace{-3 mm}
\subsubsection{Inter-Core Balancing} \label{SectionInterCoreBalancing}
%%\vspace{-1 mm}
The inter-core balancing is performed for the layers that support filter reuse because of the static nature of the weights. In the proposed dataflow, the layers that support filter reuse are the regular and the depthwise separable convolution layers. As shown in Figure \ref{conv1}, each $R\times C$ matrix column, at any given time, processes the same filter. The filters are scheduled in a \textit{low latency, more dense} and \textit{high latency, less dense} approach. Assuming that during the first iteration, the four filters (F0, F1, F2, F3) are broadcasted and the first column's processing latency is the lowest. In the next filter broadcast, the next set of filters will be broadcasted in such a way that the filter with the highest density and the associated input channel will be scheduled to the first column. In a similar fashion, as the columns proceed to completion, the next filters are scheduled based on the order of their completion. This ensures that the processing completes uniformly across all the columns so that the idle time for the individual \textit{phantom} cores is minimized. Computation of density is trivial as it involves finding the total number of ones in the sparse mask of the filters, and thus, requires minimal resources. 
%The impact of inter-core balancing on the performance will be explored in Section 5. 

\begin{figure}[ht]
\centering
\includegraphics[width=0.65\textwidth,keepaspectratio]{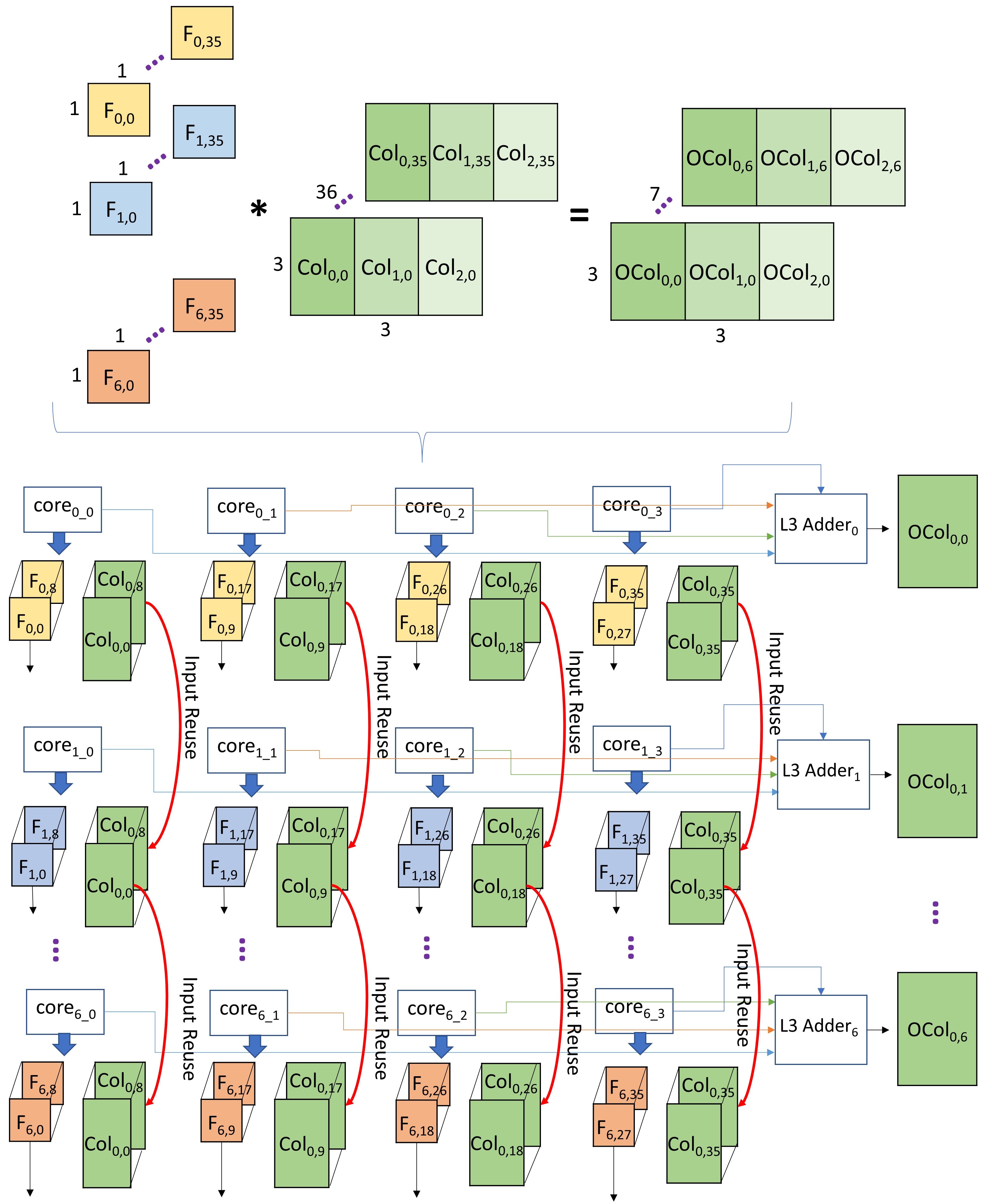}
\caption{Pointwise convolution}
\label{conv2}
\end{figure}

%%%%%%%%%%%%%%%%%%%%%%%%%%%%%%%%%%%%%%%%%%%%%%%%%%%%%%%%%%%%%%%%
%%\vspace{-2 mm}
\subsection{Pointwise Convolution}
%%\vspace{-1 mm}
%%%%%%%%%%%%%%%%%%%%%%%%%%%%%%%%%%%%%%%%%%%%%%%%%%%%%%%%%%%%%%%%
%Pointwise convolutions are quite popular in modern CNN designs. These, along with the depthwise separable, are replacing the regular 2D convolutions owing to their less number of MAC operations \cite{mobnetv1}. A pointwise convolution convolves a $1\times1\times C\times P$ filters with an $M\times N \times C$ input feature map to produce an $M \times N \times P$ output feature map. Here, $M$ and $N$ are the input width and height, respectively. $C$ is the number of input channels, and $P$ is the number of filters. 

Figure \ref{conv2} shows a pointwise convolution example where a set of $1\times 1 \times 36 \times 7$ sparse filters convolve with a $3\times3\times36$ sparse input to produce a $3\times3\times7$ output. These dimensional parameters are a good representative of the layer dimensions of actual CNNs. Figure \ref{conv2} also shows the dataflow and the mapping of the various computations involved for this example. It can be seen that the $7$ filters are scheduled along the $7$ rows of the \textit{Phantom} cores with each row processing equal number of channels along the columns. The channels are equally divided based on the number of combined multiplier threads in each \textit{Phantom} core. Since, in this design, each \textit{Phantom} core consists of 3 PEs, with each PE containing 3 multipliers (Figure \ref{CE}), the channels are divided equally into batches of 9 and scheduled along the columns to maximize the hardware utilization (G2). To enhance data reuse (G1), each weight matrix is held locally in a particular core while the input is swept across it. The input is scheduled in a \textit{channel-first} manner, followed by rows, and then columns. 
%From Figure \ref{conv2}, the input \textit{a\textsubscript{0-2\textunderscore0-8}} is a $3\times1\times9$ matrix. The 9 values \textit{a\textsubscript{0\textunderscore0-8}}, in the first row are provided first, followed by the $9$ values in the second row. 
Figure \ref{conv2} only shows the generation of the first column of the output. After all the rows for a particular channel have been exhausted, the next column is evaluated in a similar fashion till all the columns have been swept through. The L3 adder circuit gathers all the partial outputs from the cores along the columns to generate the required outputs. 
%%%%%%%%%%%%%%%%%%%%%%%%%%%%%%%%%%%%%%%%%%%%%%%%%%%%%%%%%%%%%%%%

\begin{figure}[ht]
\centering
\includegraphics[width=0.65\textwidth,keepaspectratio]{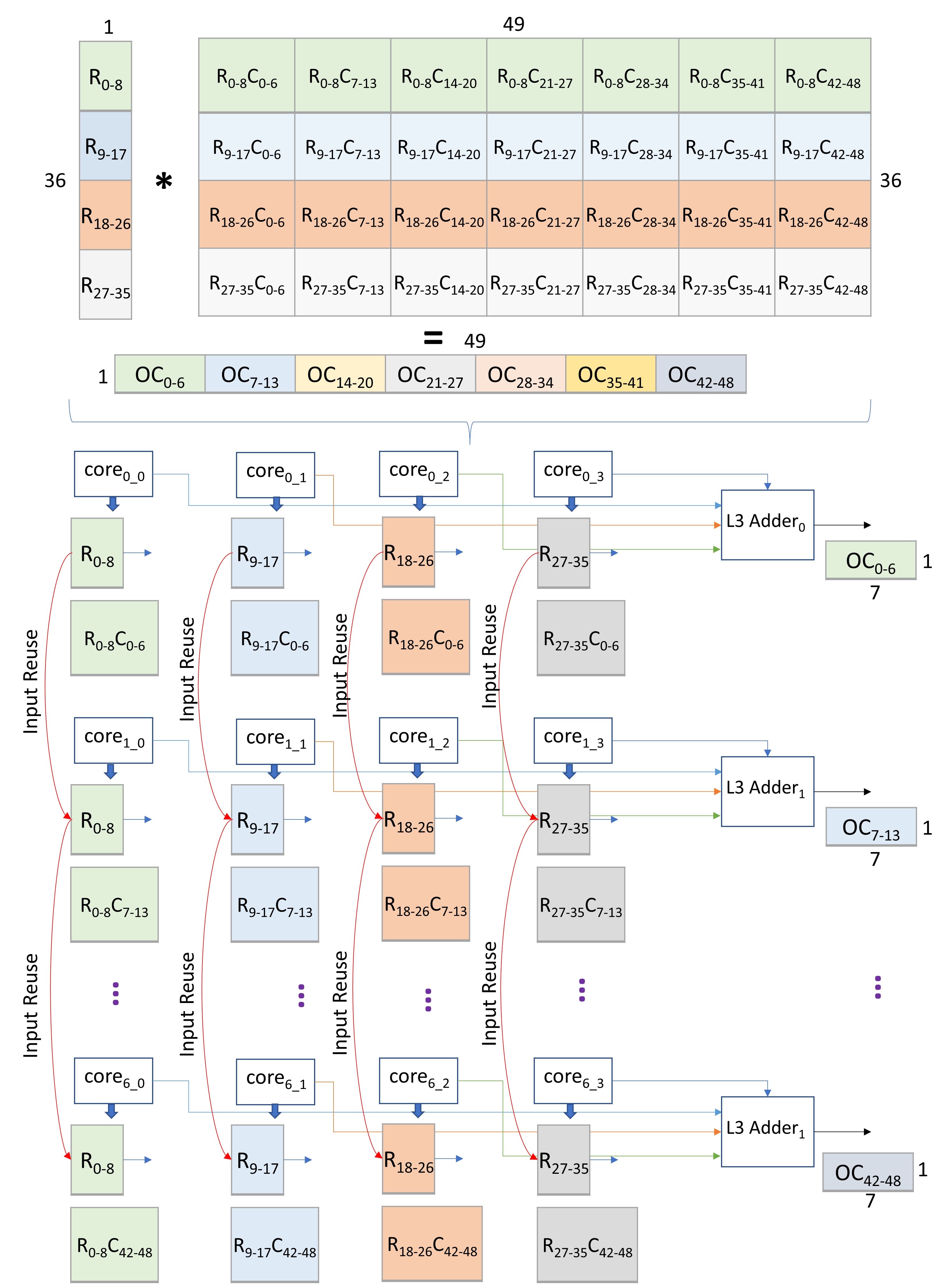}
\caption{FC layer processing}
\label{conv3}
\end{figure}

%%%%%%%%%%%%%%%%%%%%%%%%%%%%%%%%%%%%%%%%%%%%%%%%%%%%%%%%%%%%%%%%
%%\vspace{-2 mm}
\subsection{FC Layers}
%%\vspace{-1 mm}
%%%%%%%%%%%%%%%%%%%%%%%%%%%%%%%%%%%%%%%%%%%%%%%%%%%%%%%%%%%%%%%%
FC layers are an inherent part of modern CNN designs and, therefore, need to be accounted for in CNN accelerator designs (G3). AlexNet and VGG-16 both have FC layers with activation vectors that are 4K long and weight matrix that is 4K $\times$ 4K long. Similarly, Mobilenet has an FC layer with activation vector that is 1K long and weight matrix that is 1K $\times$ 1K long. The FC parameters comprise of a total of 24.3\% of all the parameters in the Mobilenet. FC computations, therefore, are crucial and need to be accounted for in CNN accelerator designs (G3). 
Figure \ref{conv3} shows an FC computation example where a length 36 sparse input vector (R\textsubscript{0}-R\textsubscript{35}) is element-wise convolved with a $36\times49$ sparse weight matrix to generate a length $49$ output vector. Figure \ref{conv3} also shows the dataflow and the computational mapping of the example onto the \textit{Phantom} cores in the \textit{Phantom-2D} architecture. Similar to the pointwise convolution, the input and the weight channels are broken into 4 batches of length 9 and scheduled across the columns. The input vector, is held stationary (input stationary) across the rows and the individual weight vectors are swept over the input vector to generate the partial outputs. Similar accumulation is performed by the L3 adders to generate the final outputs. 
%%%%%%%%%%%%%%%%%%%%%%%%%%%%%%%%%%%%%%%%%%%%%%%%%%%%%%%%%%%%%%%%

\begin{figure}[t]
\centering
\includegraphics[width=0.65\textwidth,keepaspectratio]{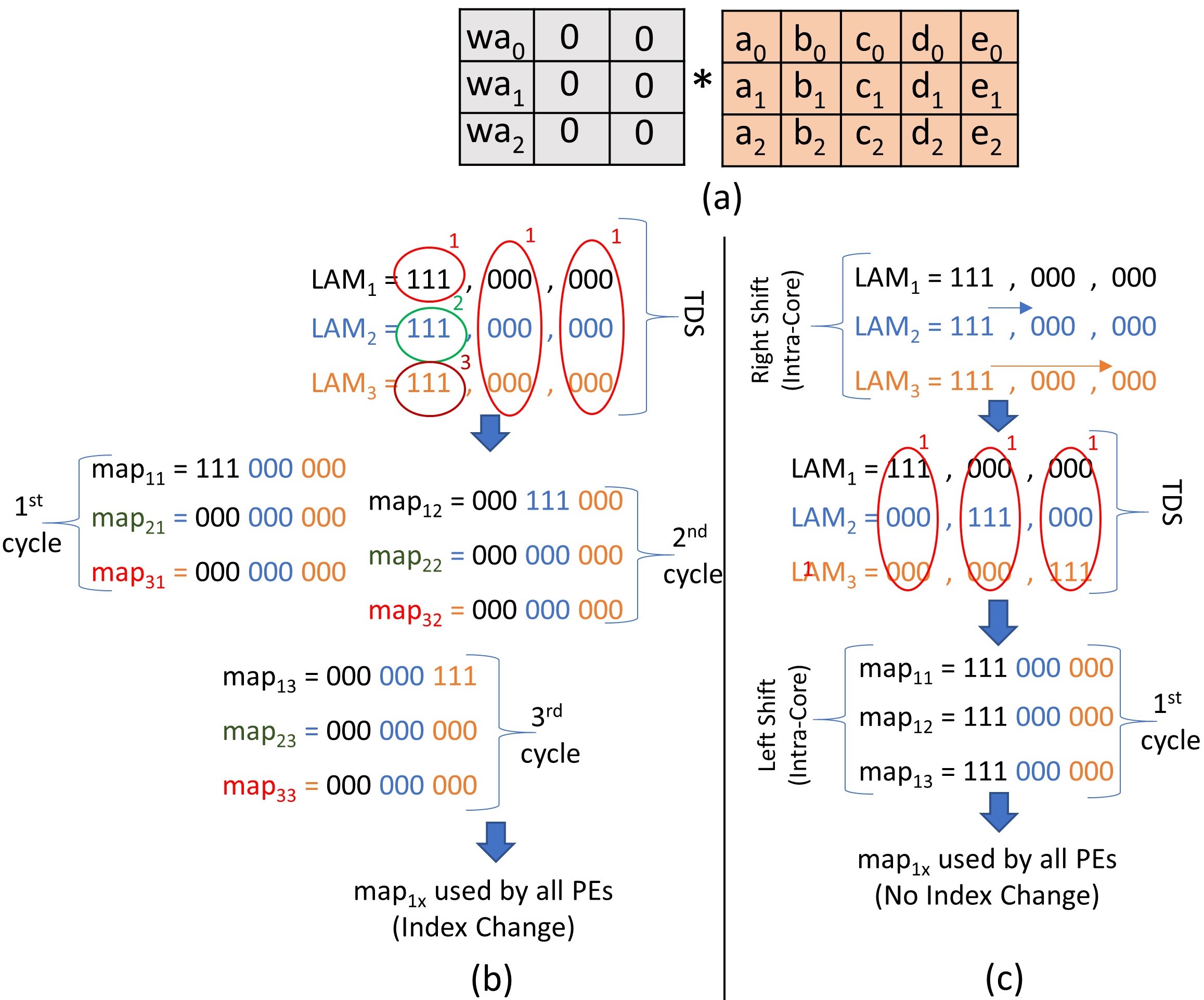}
%%\vspace{-2 mm}
\caption{Intra-core balancing (a) Test example (b) Without balancing (c) With balancing}
\label{intra}
%%\vspace{-6 mm}
\end{figure}
%%%%%%%%%%%%%%%%%%%%%%%%%%%%%%%%%%%%%%%%%%%%%%%%%%%%%%%%%%%%%%%%
%%\vspace{-2 mm}
\subsection{Intra-Core Balancing}
%%\vspace{-1 mm}
%%%%%%%%%%%%%%%%%%%%%%%%%%%%%%%%%%%%%%%%%%%%%%%%%%%%%%%%%%%%%%%%
The \textit{Intra-Core} balancer performs thread-level balancing inside every \textit{Phantom} core. This is opposed to the \textit{Inter-Core} balancer which performs balancing across all the \textit{Phantom} cores in the \textit{Phantom-2D} architecture. 
%The primary block inside the \textit{Phantom} core that determines the thread selection is the TDS (Section 3.4). 
Recall that the TDS (Section 3.4) operates in a column-wise manner where each column of the LAM outputs are evaluated concurrently, as shown in Figure \ref{TDS_1}. Also, recall that the total number of ones selected by the TDS per column are less than or equal to the number of multiplier threads per PE (3 in this case). Because of the column-wise selection, the TDS latency is bounded by the column with the highest density. Figure \ref{intra}(a) shows a test example to explain the intra-core balancing and its significance. Figure \ref{intra}(b) shows the core operation without any balancing. The first column of the weight matrix has the highest density. This is also reflected in the generated LAM values. This uneven distribution results in the first column taking the highest number of cycles, whereas, the second and the third column selection completing in only one cycle, as shown in Figure \ref{intra}(b). This stalls the core as the core must wait for the first column to complete all 3 cycles before processing the next block which significantly decreases the thread utilization of the core (Valid Computations/(Cycles $\times$ PEs $\times$ ThreadsPerPE) = 9/(3 $\times$ 3 $\times$ 3) = 33\%). \par
Figure \ref{intra}(c) shows how the intra-core balancer mitigates this issue by efficient distribution of the load, prior to the TDS operation, in a relatively simplistic manner. A right circular shift is performed on LAM\textsubscript{2} and LAM\textsubscript{3} values, as shown in Figure \ref{intra}(c). This modifies the load distribution among the three LAM columns. Operation of the TDS this time, ensures the selection of all three columns in one cycle. A circular left shift is performed on the generated map inputs (map\textsubscript{11} = 111 000 000, map\textsubscript{12} = 000 111 000, map\textsubscript{13} = 000 000 111) in the same manner as the circular right shift in the first step, to ensure a valid mapping by the mapper. The updated map values, after shifting, are shown in Figure \ref{intra}(c). The updated map values are then used in the map\textsubscript{1x} without adjustment in the location/index (Section 3.5) to accurately map data to the individual threads. The thread utilization of the core, in this case, increases drastically (Valid Computations/(Cycles $\times$ PEs $\times$ ThreadsPerPE) = 9/(1 $\times$ 3 $\times$ 3) = 100\%). Finally, it should be noted that the evaluation of the example in Figure \ref{intra}, is performed with $L_f=3$. However, the same process follows for any value of $L_f$. We will further explore the combined affect of both, the inter-core, and the intra-core balancing, on performance, in Section 5.
%

%%%%%%%%%%%%%%%%%%%%%%%%%%%%%%%%%%%%%%%%%%%%%%%%%%%%%%%%%%%%%%%%
%%\vspace{-3 mm}
\section{Experiments and Results}
%%\vspace{-3 mm}
\subsection{Evaluation Methodology}
%%\vspace{-1 mm}
%%%%%%%%%%%%%%%%%%%%%%%%%%%%%%%%%%%%%%%%%%%%%%%%%%%%%%%%%%%%%%%%
\noindent\textbf{Cycle-Accurate Simulator:} To accurately model the performance of an individual \textit{Phantom} core and the \textit{Phantom-2D} accelerator, as a whole, we design a software-based, cycle-level performance simulator. The simulator is parameterizable across various core and architecture level design parameters to capture their effect on the performance. Table \ref{params} shows the modifiable design parameters. The operations are categorized as core level and architecture level. Table \ref{config} shows the arrangement and configuration of the \textit{Phantom-2D} accelerator's compute matrix containing \textit{Phantom} cores. 
%It has a $7\times4$ matrix of \textit{Phantom} cores with each core containing a total of $9$ threads, distributed equally among the three PEs in a multi-threaded design. 

\begin{table}[t]
  \centering
   \caption{Operation parameters }
  \label{params}
  \begin{tabular}{|c|c|c|}
    \hline
    \textbf{Operation} & \textbf{Level} & \textbf{Parameters}\\
    \hline
    TDS  & \textit{Phantom} & TDS\textunderscore inOrder\\ 
    & & TDS\textunderscore outOrder\\
    
    \hline
    Balancing  &  & unbalanced\\ 
    &  \textit{Phantom}/ & intra\textunderscore core\\ 
    & \textit{Phantom-2D} & inter\textunderscore core\\
    & & full(inter + intra)\\
    
    \hline
    {Lookahead factor ($L_f$)}  & \textit{Phantom} & $3\leq L_f \leq 27$\\
    
     \hline
    CNN models  & \textit{Phantom-2D} & dense\\ 
    & & sparse\\
    \hline
  \end{tabular}

\end{table}

\begin{table}[t]
 %%\vspace{-3 mm}
  \centering
  \caption{Accelerator configuration }
  \label{config}
  \begin{tabular}{|c|c|}
    \hline
    \textbf{Configuration parameter} & \textbf{Value}\\
    \hline
    Compute matrix size  & 28 ($7\times4$)\\
    
    \hline
    PEs per core  & 3\\ 
    
    \hline
    Multiplier threads per PE  & 3\\
    
     \hline
    Total multiplier threads & 252\\ 
    
    \hline
  \end{tabular}
\end{table}

%%%\vspace{-1 mm}

The simulator has a set of 5 built-in test scenarios which covers the sweeping of all the parameters shown in Table \ref{params}. The simulator also contains routines for SparTen, SCNN, and Eyeriss v2 for performing comparisons. During the processing of each layer, every \textit{Phantom} core outputs 7 values which includes the total cycles by a dense design, the cycles of TDS in-order and TDS out-of-order, coupled with two-level load balancing, and the average thread utilization for various \textit{Phantom-2D} configurations. The total cycle count for SparTen, SCNN, and Eyeriss v2 is also outputted. The evaluation files in the simulator use the data provided by the individual routines and schedulers to generate the throughput and speedup results for dense and sparse accelerator designs.\\
\textbf{Simulated Models:} Although we test our design on many CNN models including Alexnet \cite{alexnet}, VGG16 \cite{vgg16}, MobileNets \cite{mobnetv1,mobnetv2}, GoogleNet \cite{googlenet}, and recently proposed EfficientNetV2 \cite{enet}, we present the results for sparse versions of VGG16 and MobileNet for comparison purposes.
%The reason for our selection of these models is because they cover the different types of layers usually present in all CNN models. These layers include regular and depthwise separable, unit and non-unit stride convolutions and FC layers. 
We use the approach presented in \cite{deepcompression} to prune these networks using the MATLAB's \textit{Deep Learning Toolbox}, and ensure that we achieve the same level of weight sparsity as previous approaches, for fair comparisons. The activation sparsity is highly dependent on the input and changes dynamically during the inference process. We, therefore, average out the input sparsity for a batch of 100 randomly selected inputs. After pruning of the network, we generate the sparse binary masks for every layer and generate a network containing only the sparse masks, since only this information is needed to efficiently represent the MAC operations needed per layer for the \textit{Phantom-2D} accelerator. The sparse masks are fed into the simulator in the model dimensions (i.e., mapped to the dimension of each CNN layer in the model). The simulator's scheduler uses the dataflow for various layers, presented in Figures \ref{conv1}-\ref{conv3}, to break down the dimensions, and schedule the individual binary masks to different \textit{Phantom} cores.\\
%\textbf{FPGA Implementation:} We write RTL Verilog for the \textit{Phantom} cores and implement our design on Xilinx Z-7100 SoC. The SoC consists of a programmable logic (PL) and a processing system (PS) part. The PS acts as a CPU and transfers layer data from the DRAM to the PL on which our RTL is implemented. The PL has over 554K and 277K LUTs and flip-flops, respectively. It has a total of 26.5 Mb of on-chip SRAM capacity. The design implementation also includes LAM blocks consisting of AND gates. The TDS, Mapper, CE, and OB implementations are shown in Figures \ref{TDS_3}, \ref{map}, \ref{CE}, and \ref{OB}, respectively. The design is synthesized using Xilinx Vivado 2020 synthesis tool.

%%%%%%%%%%%%%%%%%%%%%%%%%%%%%%%%%%%%%%%%%%%%%%%%%%%%%%%%%%%%%%%%

\begin{figure}
\centering
\includegraphics[width=0.70\textwidth,keepaspectratio]{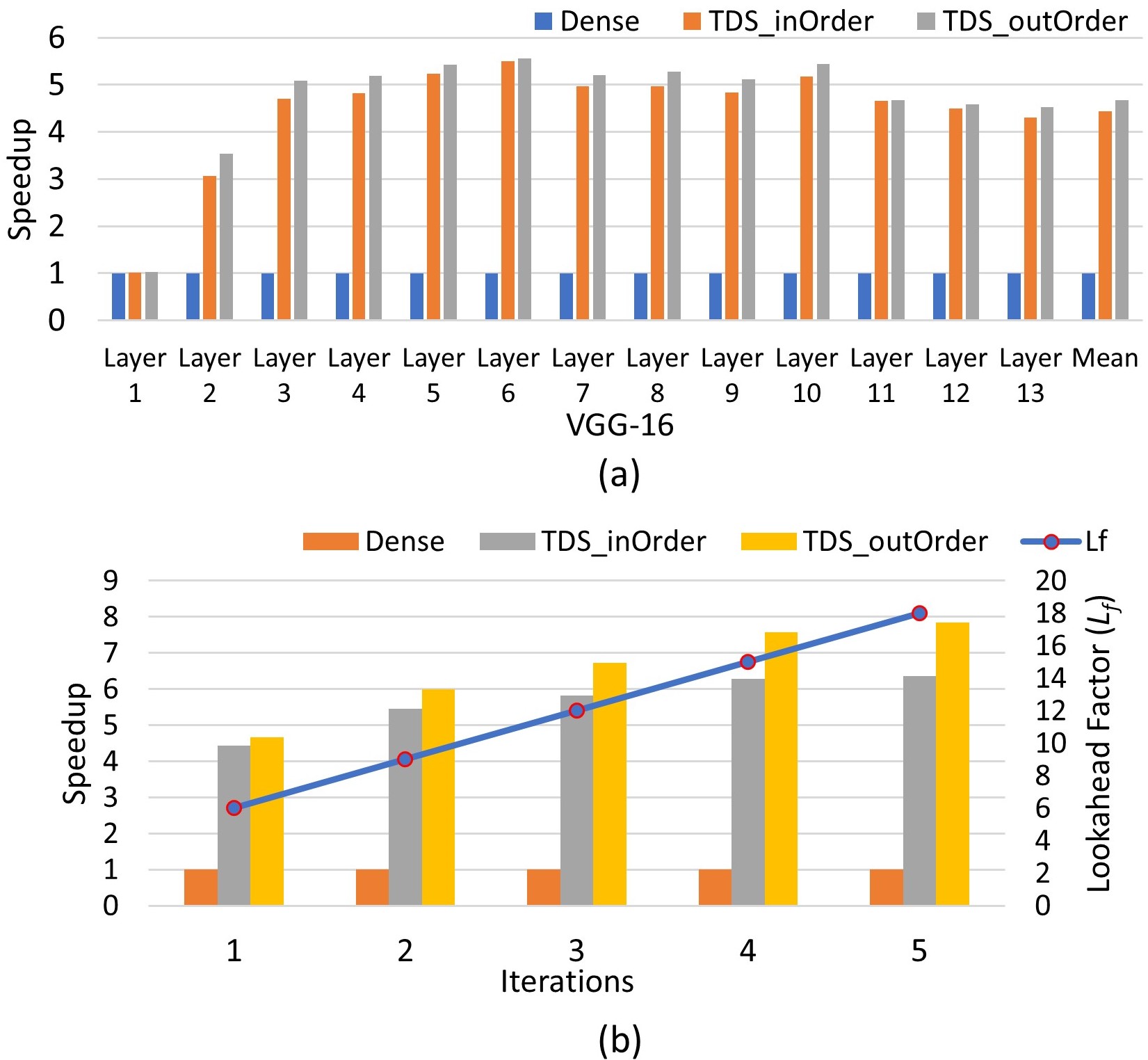}
%%\vspace{-4 mm}
\caption{TDS-IO vs. TDS-OO (a) Per layer comparison with $L_f=6$ (b) Comparisons with changing $L_f$}
\label{test1}
%%\vspace{-7.0 mm}
\end{figure}
%%%%%%%%%%%%%%%%%%%%%%%%%%%%%%%%%%%%%%%%%%%%%%%%%%%%%%%%%%%%%%%%
%%\vspace{-2 mm}
\subsection{Results}
%%\vspace{-1 mm}
%%%%%%%%%%%%%%%%%%%%%%%%%%%%%%%%%%%%%%%%%%%%%%%%%%%%%%%%%%%%%%%%
We capture the simulator's results in an incremental approach, starting from the basic core-level configurations, and moving towards system-level configurations. All the comparisons are made with an equivalent dense architecture, having an equal number of MAC units, but without leveraging the sparse optimizations. At the end, we compare different versions of \textit{Phantom-2D} accelerator against previously proposed two-sided sparse architectures.

%%\vspace{-2 mm}
\subsubsection{TDS Variants Comparison}
%%\vspace{-1 mm}
Figure \ref{test1}(a) shows the performance comparison of the two TDS variants, namely, the in-order TDS (TDS-IO) and the out-of-order TDS (TDS-OO), evaluated on a sparse VGG16 net. We set the lookahead factor ($L_f$) for this test to 6 for both the variants. The performance of the dense architecture can be modeled by setting the $L_f$ value equal to 1. This would ensure that no future computations are observed by the TDS, thus, replicating a dense accelerator. It can be seen that the first layer does not have any significant performance gains over a dense architecture because of very low sparsity. The performance gains increase drastically in subsequent layers with TDS-IO variant, on average, being $4.5\times$, and the TDS-OO variant, being $4.8\times$ faster, than the equivalent dense architecture. This represents a $1.07\times$ performance improvement of TDS-OO over TDS-IO. The performance difference, however, improves substantially as $L_f$ is increased, as shown in Figure \ref{test1}(b). We run the VGG16 net a total of five times, and average out the speedups, while sweeping $L_f$ from 6 to 18, with a jump of three in every iteration. For $L_f$ = 18, we observe a $6.35\times$ and a $7.9\times$ performance gain of TDS-IO and TDS-OO, respectively, over dense architecture, representing a $43\%$ and a $68\%$ increase, when switching from $L_f$ = 3 to $L_f$ = 18, respectively. TDS-OO, for $L_f$ = 18, gives a $1.24\times$ performance gain over TDS-IO, a jump of $16\%$ from $L_f$ = 6. These results are inline with our preliminary observations in Section 3.4. The improvement in the core's thread utilization from TDS-OO directly correlates to the increase in performance. For the next experiments, we will always use the TDS-OO variant.

\begin{figure}
\centering
\includegraphics[width=0.75\textwidth,keepaspectratio]{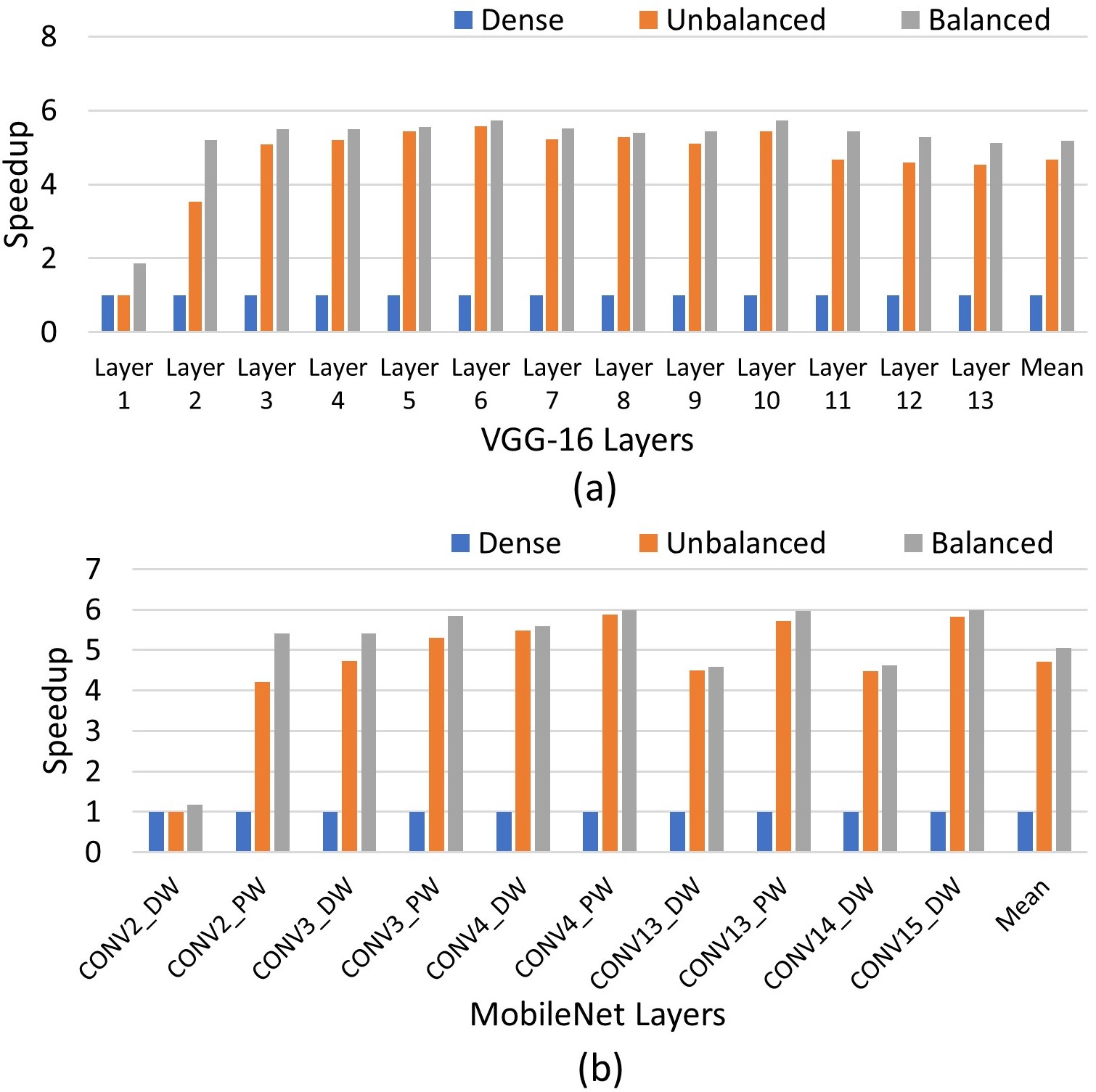}
%%\vspace{-4 mm}
\caption{Balanced vs unbalanced at $L_f=6$ for (a) Sparse VGG16 (b) Sparse MobileNet}
\label{test2}
%%\vspace{-6.5 mm}
\end{figure}

%%\vspace{-3 mm}
\subsubsection{Impact of Load Balancing}
%%\vspace{-1 mm}
Figure \ref{test2} shows the impact of load balancing (\textit{inter-core} + \textit{intra-core}) on the performance for sparse VGG16 and MobileNet with $L_f=6$. During the initial layers for both the CNNs, we observe a drastic improvement in performance (as much as $1.5\times$ for VGG16 and $1.3\times$ for MobileNet). On average, we observe a performance difference of $1.1\times$ and $1.08\times$ for VGG16 and MobileNet, respectively. From our experiments, we also observe that the performance difference due to \textit{intra-core} balancing increases drastically at high sparsity and greater value of $L_f$. This conclusion is inline with the example in Figure \ref{intra}, where the increase in thread utilization from \textit{intra-core} balancing increases the speedup by $3\times$. The \textit{inter-core} balancing is dominant in later layers where there are a large number of channels. For reduction in simulation times, we only use approximately $25\%$ of the channel filters for our simulations in the case of regular (and depthwise seperable) convolutions, which prevents us from exploiting the full power of the \textit{inter-core} balancing. Hence, Figures \ref{test2}(a) and (b) do not show a significant improvement in layers with a large number of channels. Based on our experiments, we observe, on average, a $7\%$ increase in speedup by having 15\% more filters in our simulations.

%%\vspace{-2 mm}
\subsubsection{Sensitivity to Sparsity and $L_f$}
\begin{figure}[t]
\centering
\includegraphics[width=0.80\textwidth,keepaspectratio]{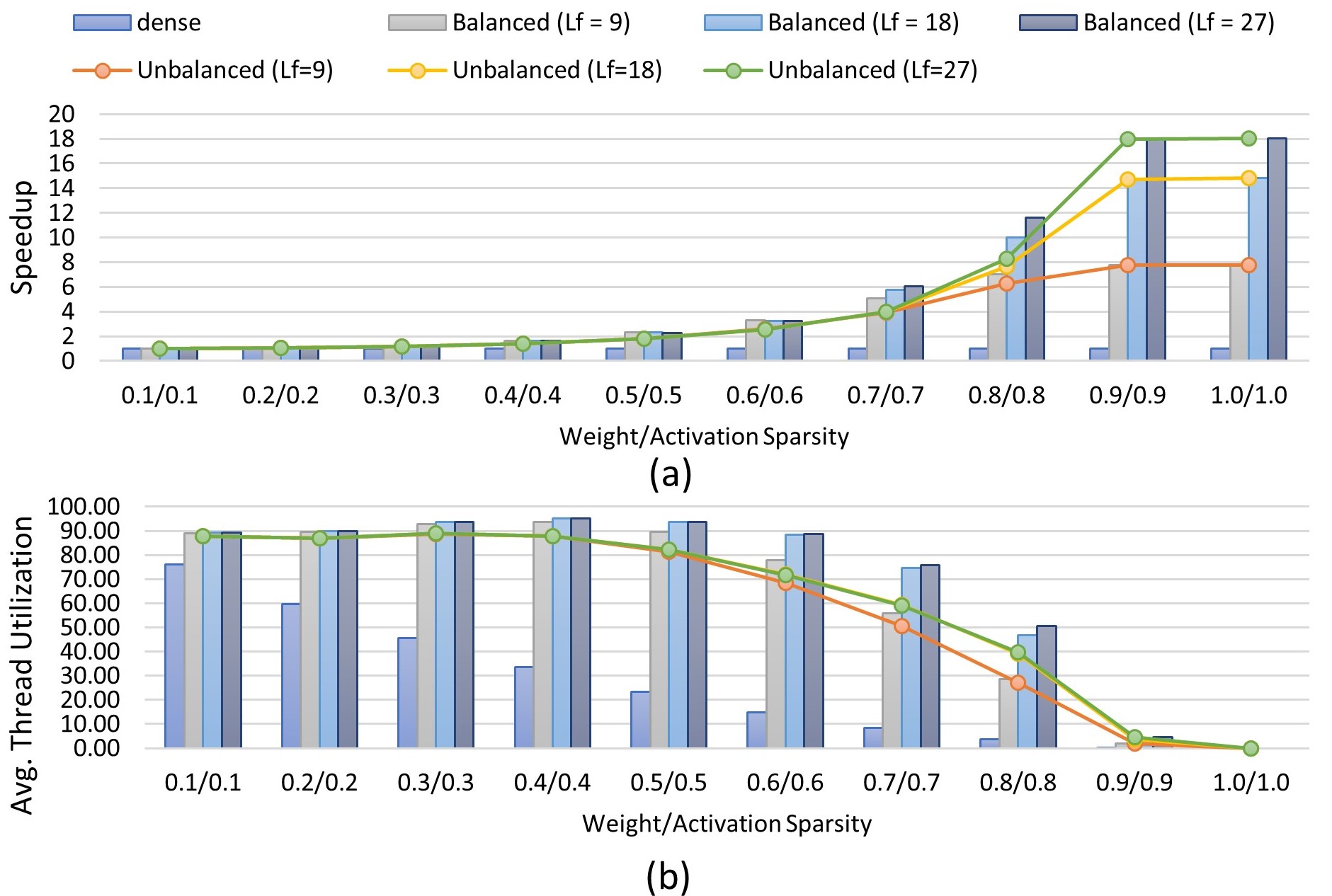}
%%\vspace{-4 mm}
\caption{Sensitivity to sparsity and $L_f$ for VGG16 (a) Speedup (b) Average thread utilization}
\label{test4_a}
%%\vspace{-7.0 mm}
\end{figure}

\begin{figure}
\centering
\includegraphics[width=0.80\textwidth,keepaspectratio]{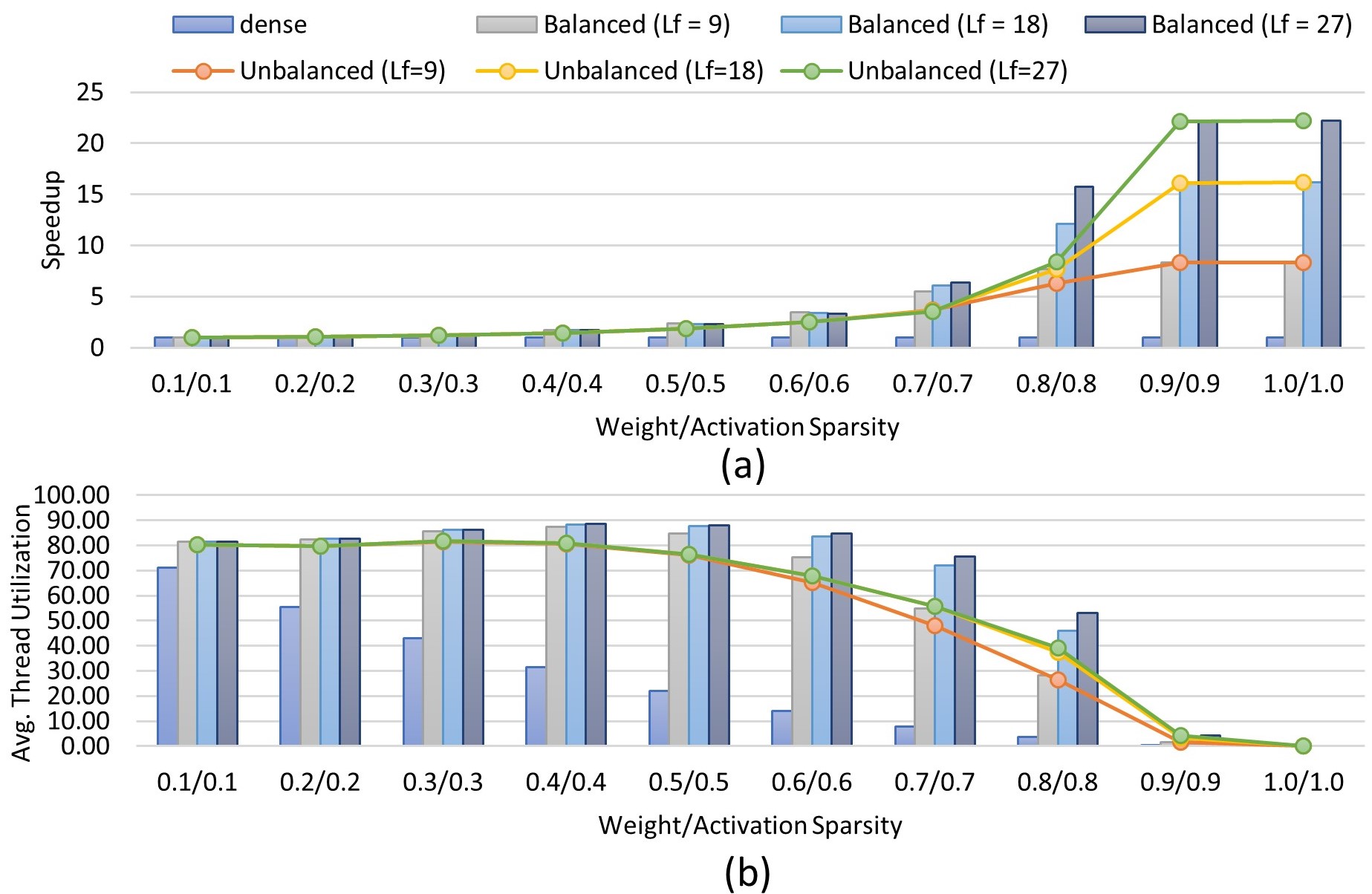}
%\vspace{-4 mm}
\caption{Sensitivity to sparsity and $L_f$ for MobileNet (a) Speedup (b) Average thread utilization}
\label{test4_b}
%\vspace{-7 mm}
\end{figure}

%%\vspace{-1 mm}
Our performance simulator has the capability to sweep the weight/activation sparsity from low 0.1/0.1 (10\%) to high 1.0/1.0 (100\%). This is shown in Figures \ref{test4_a} and \ref{test4_b}. We also plot the average multiplier thread utilization at different levels of sparsity for VGG16 and MobileNet. For a dense architecture, we observe a higher thread utilization at low sparsity and lower thread utilization at a higher sparsity, as shown in Figures \ref{test4_a}(b) and \ref{test4_b}(b). \textit{Phantom-2D}, however, exhibits significantly higher thread utilization compared to a dense architecture, even at high levels of sparsity. For VGG16, \textit{Phantom-2D} consistently keeps the thread utilization higher than $90\%$ even at 60\% sparsity for both weights and activations, whereas, as expected, the utilization for the dense architecture decreases by $25-30\%$ and decreases by almost 50\% at higher sparsity levels. This thread utilization difference directly correlates to greater speedups at mid to high levels of sparsity, as shown in Figures \ref{test4_a}(a) and \ref{test4_b}(a). At higher sparsity levels (0.8/0.8 - 1.0/1.0), we observe a massive speedup over dense architecture even when the thread utilization starts decreasing. This is because the \textit{Phantom-2D} accelerator starts actively skipping all the \textit{zero} computations without wasting compute cycles.\par

For convenience in comparing different versions of the \textit{Phantom-2D} accelerators, we rename them as \textit{Phantom-2D-CV} (conservative, with $L_f = 9$, balanced), \textit{Phantom-2D-MD} (moderate, with $L_f = 18$, balanced), and \textit{Phantom-2D-HP} (high-performance, with $L_f = 27$, balanced). At low sparsity, all three versions exhibit similar speedups and thread utilization. This difference increases at higher sparsity levels with \textit{Phantom-2D-MD} and \textit{Phantom-2D-HP} being $1.43\times$ and $1.65\times$ faster, respectively, than \textit{Phantom-2D-CV}, at $80\%$ sparsity. The unbalanced configurations display a similar trend among each other. Comparison between the \textit{Phantom-2D-HP}, balanced and unbalanced configurations, show a $1.4\times$ speedup of balanced over unbalanced at 80\% sparsity. Analysis of hardware resources show that the \textit{Phantom-2D-HP} requires only $1.05\times$ more LUTs than \textit{Phantom-2D-CV}. This is because the higher values of $L_f$ only increase the LUT count of LAM and TDS blocks, whereas, the Mapper, CE, and the OB blocks' LUT count remains the same.

%%\vspace{-2 mm}
\subsubsection{Comparison Against Past Approaches}
%%\vspace{-1 mm}

\begin{figure*}
\centering
\includegraphics[width=0.80\textwidth,keepaspectratio]{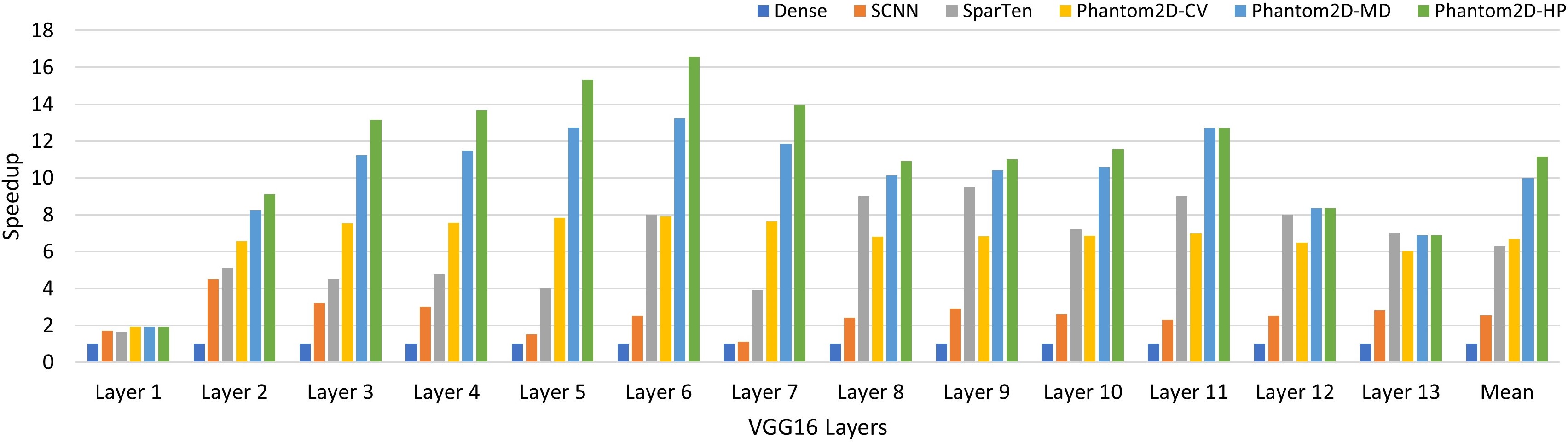}
%\vspace{-4 mm}
\caption{VGG16 speedup comparison with SCNN\cite{scnn} and SparTen\cite{sparten}}
\label{test5_a}
%\vspace{-6 mm}
\end{figure*}

Figure \ref{test5_a} shows the speedup comparison of the different versions of the \textit{Phantom-2D} accelerators over dense architecture, SCNN, and SparTen, for sparse VGG16 net. The average sparsity for the weights and activations is 77\% and 68\%, respectively. Note that both the SCNN, and SparTen, do not support FC layers, whereas, \textit{Phantom-2D} does. In addition, SCNN also does not support non-unit stride convolutions present in Alexnet and MobileNet. Therefore, for fair comparison, we omit the last three FC layers in our results and use sparse VGG16 which does not contain any non-unit stride convolution. 
%The speedup values for SCNN and SparTen are extracted from Figure 9 in SparTen\cite{sparten}.
We observe that the \textit{Phantom-2D-CV} version, on average, performs $1.05\times$, $2.56\times$, and $6.4\times$, better than SparTen, SCNN, and dense architecture, respectively. The speedup increases to approximately $1.57\times$, $3.8\times$, and $9.9\times$ for \textit{Phantom-2D-MD}, and $1.98\times$, $4.1\times$, and $11\times$, for \textit{Phantom-2D-HP} over SparTen, SCNN, and dense architecture, respectively. Upon comparing different versions of \textit{Phantom-2D}, we observe that, on average, \textit{Phantom-2D-HP} has a $67\%$ and a $14\%$ improvement in performance, compared to the \textit{Phantom-2D-CV} and \textit{Phantom-2D-MD} versions, respectively. 
%Looking deeper into layer by layer comparisons, we observe that the \textit{Phantom-2D-HP} (plus \textit{CV} and \textit{MD}) offer huge speedups in the initial layers over SparTen (and SCNN) because of \textit{Phantom's} intraCore balancing. However, layers 8, 11, 12, and 13 show a decrease in speedup. This follows from our previous discussions related to not exploiting the full power of interCore balancing. SparTen, however, exploits this balancing using their \textit{greedy balancing} approach which exceeds their speedup for layers having small input dimensions but higher channel count. 
Lastly, we would like to point out the impact of FC layers on the \textit{Phantom-2D} versions. Our experiments show, on average, a speedup of $13\times$, $11.4\times$, and $8.6\times$ for \textit{Phantom-2D-HP}, \textit{Phantom-2D-MD}, and \textit{Phantom-2D-CV}, respectively, over dense architecture after inclusion of FC14, FC15, and FC16 layers of the VGG16 net. This improvement corresponds to our efficient dataflow and scheduling for the FC layers. \par

\begin{figure}
\centering
\includegraphics[width=0.80\textwidth,keepaspectratio]{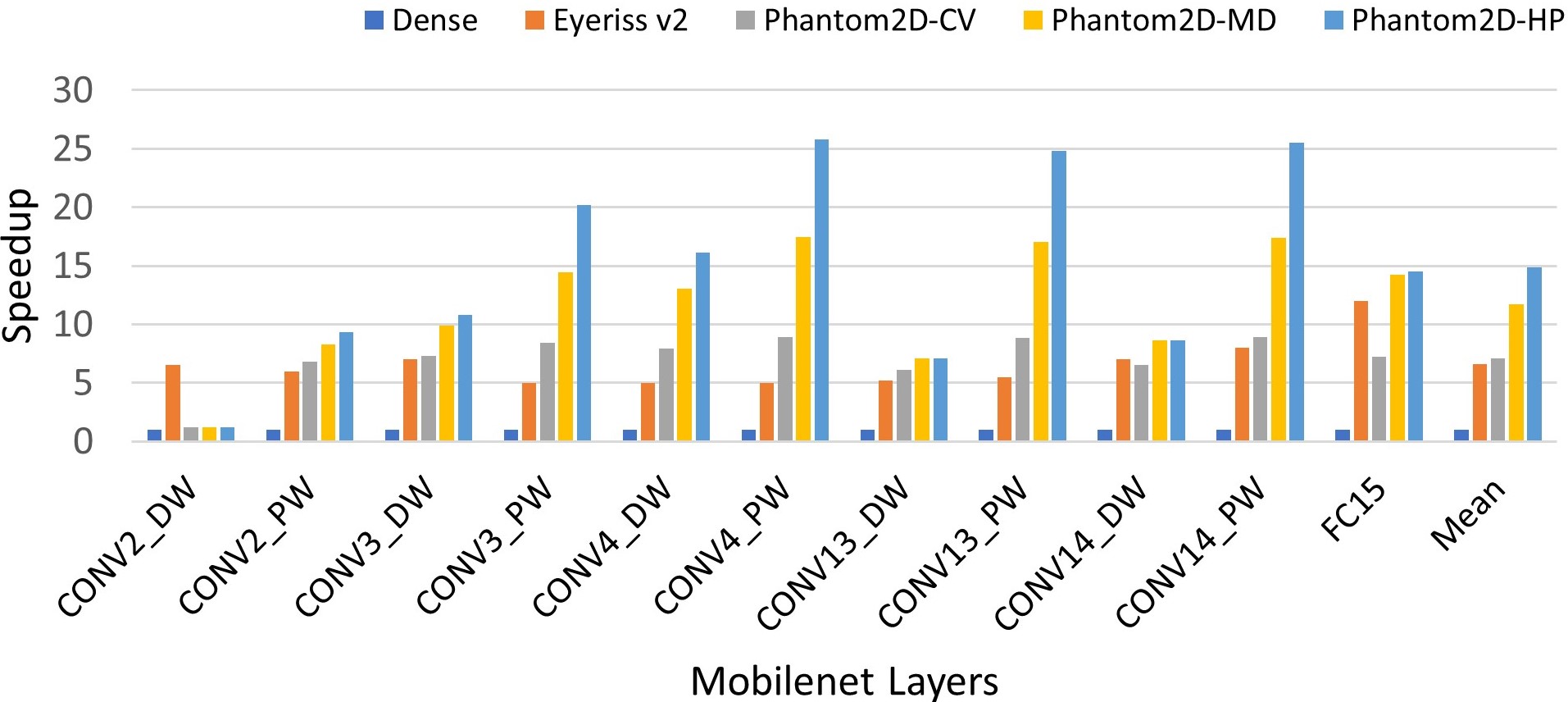}
%\vspace{-4 mm}
\caption{Speedup comparison with Eyeriss v2\cite{eyerissv2}}
\label{test5_b}
%\vspace{-7 mm}
\end{figure}
Figure \ref{test5_b} shows the performance comparison of the \textit{Phantom-2D} versions against Eyeriss v2 on sparse MobileNet. The average sparsity for the weights and activations is 73\% and 64\%, respectively. 
%We extracted the Eyeriss v2 results from Figure 21(a) in \cite{eyerissv2}. 
Authors of Eyeriss v2 performed their comparisons against their previous approach (Eyeriss\cite{eyeriss}). Eyeriss v2 uses twice the number of MACs as Eyeriss, which doubles their final speedups. For fair comparison, we adjust the speedup values of Eyeriss v2 for comparison against a dense accelerator having the same number of MACs and only use the layers used by Eyeriss v2. We observe that, on average, \textit{Phantom-2D-CV} performs $1.04\times$ better than Eyeriss v2, whereas, the \textit{MD} and \textit{HP} versions, on average perform $1.71\times$ and $2.86\times$, respectively, better than Eyeriss v2. Doing a layer by layer comparison, we observe that Eyeriss v2 performs better than \textit{Phantom} in CONV2-DW (DW indicates depthwise convolution) because of its efficient hierarchical mesh network-on-chip (NoC), however, \textit{Phantom-2D-HP} catches up in deeper layers and almost always provides an improvement. Pointwise layers are especially faster in the case of \textit{Phantom-2D-HP}, with the average speedup of $4.5\times$ over Eyeriss v2, and $25\times$ over a dense architecture. This is because of the efficient channel wise breakdown offered by our dataflow for these layers. 
%Eyeriss v2 offers a slightly higher speedup in the FC layer, compared to both \textit{Phantom-2D-HP} and \textit{Phantom-2D-MD}, with Eyeriss v2 being approximately $3.44\%$ faster. 

\begin{figure}
\centering
\includegraphics[width=0.65\textwidth,keepaspectratio]{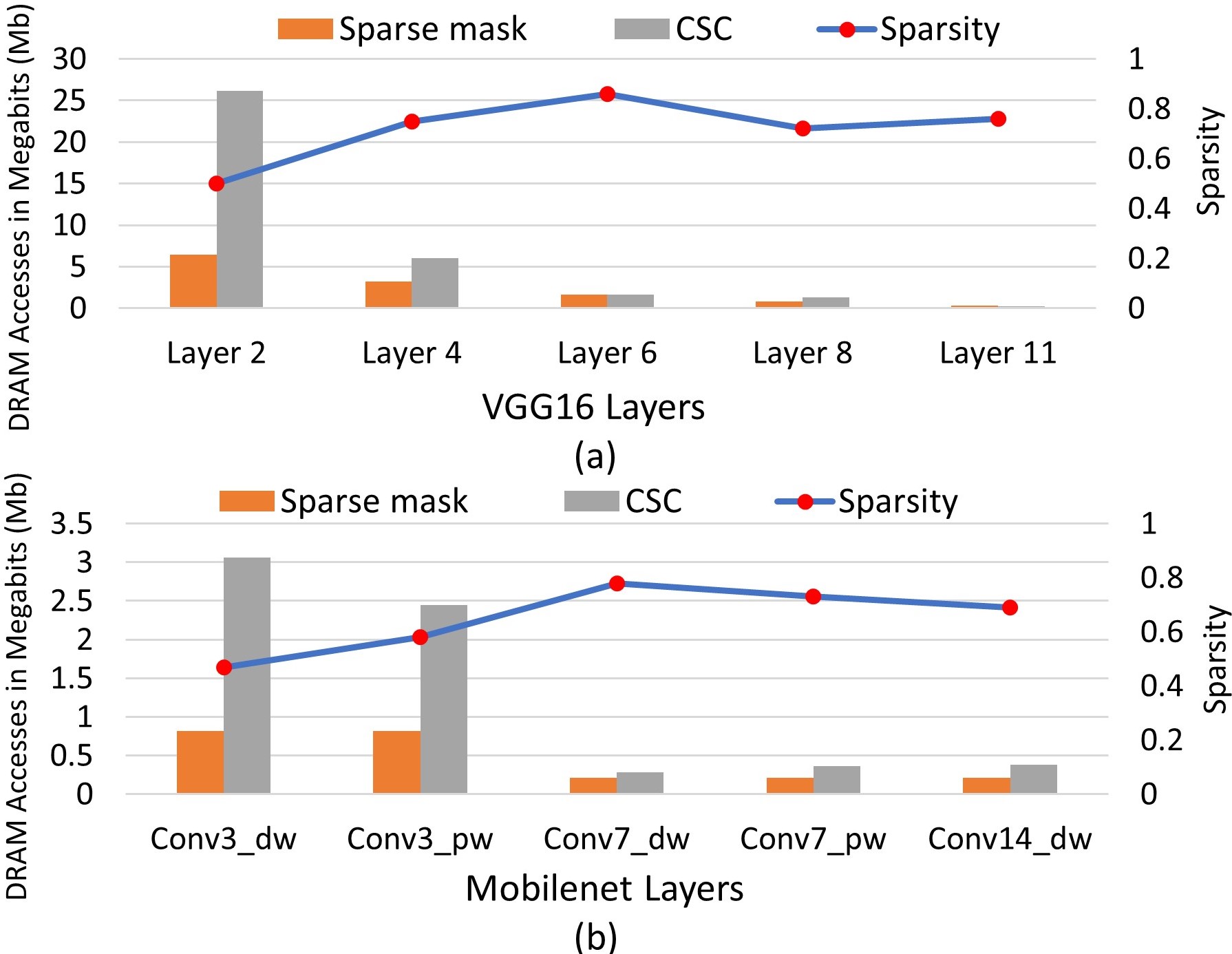}
\caption{Sparse mask vs. CSC DRAM access for (a) VGG16 (b) MobileNet}
\label{mask_csc}
\end{figure}

Comparing different versions of the \textit{Phantom-2D} accelerators, \textit{Phantom-2D-HP}, on average, offers $108.9\%$ and $27.4\%$ increase in speedup over \textit{Phantom-2D-CV} and \textit{Phantom-2D-MD}, respectively, for sparse MobileNet.\par
Comparing energy among different accelerators is quite challenging as it requires working RTLs. The energy consumption of an accelerator is dominated by the DRAM accesses, therefore, by estimating the DRAM accesses, energy difference among the accelerators can be approximated. Since many of the recent works rely on the CSC format for non-zero data storage, we compare the accessed memory for the CSC format against the sparse mask format. Figure \ref{mask_csc} shows the intermediate activations' memory access comparison for selected VGG 16 and MobileNet layers \footnote{The required memory for the stored non-zero data is not shown since it is the same for both the sparse mask and the CSC format. The accessed memory is shown for the binary sparse mask and the location vectors (column, index) of the CSC format.}. The activation sparsity for different layers is also shown. In the initial layers with low activation sparsity, the CSC format has approximately $4\times$ and $3.7\times$ higher DRAM memory accesses than the sparse mask for VGG16 and Mobilenet, respectively. In the deeper layers with moderate to high sparsity, the memory requirement for the CSC format is around $1.7\times$ that of the sparse mask.\par 
This shows that the sparse mask representation not only needs less encoding/decoding logic, but is also efficient in terms of memory requirements when compared against the CSC format. This translates directly to higher energy, area, and compute savings for the accelerators employing the sparse mask format.

%%%%%%%%%%%%%%%%%%%%%%%%%%%%%%%%%%%%%%%%%%%%%%%%%%%%%%%%%%%%%%%%

%%%%%%%%%%%%%%%%%%%%%%%%%%%%%%%%%%%%%%%%%%%%%%%%%%%%%%%%%%%%%%%%%%%%%%%%%
\subsubsection{RTL Synthesis Results}
%%%%%%%%%%%%%%%%%%%%%%%%%%%%%%%%%%%%%%%%%%%%%%%%%%%%%%%%%%%%%%%%%%%%%%%%%
We wrote the RTL Verilog for a single \textit{Phantom} core with $L_f=27$ (high performance) and synthesized it for the Xilinx Zynq 7100 SoC's programmable logic (PL), running at 150 MHz. The SoC's ARM-based processing system (PS) was used to transfer data to/from a desktop computer to DRAM. The PL acquires this data and stores it in its global buffers (input and weight SRAMs), shown in Figure \ref{ph3d}. The generated outputs and the sparse masks are stored in the output SRAMs and transferred to DRAM for the next layer. Table \ref{table1} shows the total resource utilization among various sub-blocks of the \textit{Phantom} core. The main takeaway is that the novel components of the \textit{Phantom} core (LAM, TDS, Mapper, and intra-core balancer) only account for approximately 48\% and 35\% of the utilized LUT and the FF cost, respectively. The local SRAM utilization is dominated by the mapper and the output buffering block (approximately 78\%). The design has a power consumption of 2.48W with the PS dominating the consumption (55\%). Eyeriss v2, implemented on a 65nm ASIC, utilizes 2695k gates which represents a 108\% increase in area cost when compared to the original semi-sparse Eyeriss \cite{eyeriss}. This drastic increase in area is the result of the complex encoding/decoding logic required by the CSC format in their design. SCNN has a 35\% increase in area compared to their dense design, whereas, SparTen does not report the resource utilization of their PEs.

\begin{table}
\centering
\caption{Resource utilization for a single \textit{Phantom} core with $L_f=27$}
\label{table1}
\begin{tabular}{|c|c|c|c|}
\hline
Property& Available & Used & Utilization (\%) \\
\hline
LUTs & 277k & 3.4k & $1.23\%$ \\
FFs & 554k & 6k & $1.1\%$ \\
On-chip SRAM & 26.5Mb & 2.1kB & $0.01\%$ \\
%Power & 2.48 W & NA & NA \\
\hline
\end{tabular}
\end{table}

%%%%%%%%%%%%%%%%%%%%%%%%%%%%%%%%%%%%%%%%%%%%%%%%%%%%%%%%%%%%%%%%

%%%%%%%%%%%%%%%%%%%%%%%%%%%%%%%%%%%%%%%%%%%%%%%%%%%%%%%%%%%%%%%%

%%%%%%%%%%%%%%%%%%%%%%%%%%%%%%%%%%%%%%%%%%%%%%%%%%%%%%%%%%%%%%%%
%\vspace{-3.0 mm}
\section{Conclusion}
%\vspace{-1 mm}
%%%%%%%%%%%%%%%%%%%%%%%%%%%%%%%%%%%%%%%%%%%%%%%%%%%%%%%%%%%%%%%%
Designing a CNN accelerator to leverage the two-sided sparsity is quite challenging owing to the varying layer shapes and sizes, associated with a sparse model. In this paper, we have introduced \textit{Phantom}: a novel \textit{multi-threaded}, flexible, neural computational core that exploits the two-sided sparsity to provide high gains in performance at a relatively low hardware complexity. Using a system of \textit{Phantom} cores, we then design the \textit{Phantom-2D} accelerator, and present a novel dataflow that efficiently uses the capabilities of the \textit{Phantom} cores. As opposed to many previous approaches, the \textit{Phantom-2D} accelerator supports \textbf{all} layers of a CNN, including unit and non-unit stride convolutions, and FC layers. In addition, we have also proposed a two-level load balancing strategy that efficiently balances the load across the architecture level (\textit{inter-core}), and at the thread level (\textit{intra-core}) to minimize idling of the compute threads, thereby, further increasing the throughput. We have also performed comparisons against many previous state-of-the-art two-sided  sparse CNN accelerators. Our simulations show that, on average, \textit{Phantom-2D} accelerator performs $12\times$, $4.1\times$, $1.98\times$, and $2.36\times$, better than an equivalent dense CNN architecture, SCNN, SparTen, and Eyeriss v2, respectively, while retaining the energy efficiency of SparTen.

%%%%%%%%%%%%%%%%%%%%%%%%%%%%%%%%%%%%%%%%%%%%%%%%%%%%%%%%%%%%%%%%

\bibliographystyle{ACM-Reference-Format}
\bibliography{sample-base}

\end{document}